\documentclass[12pt,preprint]{aastex}
\usepackage{emulateapj5,apjfonts}
\usepackage{epsfig}
\usepackage{amsmath}
\usepackage{amssymb}
\usepackage{onecolfloat}

\newcommand{\be}{\begin{equation}}
\newcommand{\ee}{\end{equation}}
\newcommand{\ba}{\begin{eqnarray}}
\newcommand{\ea}{\end{eqnarray}}
\newcommand{\brr}{\begin{array}}
\newcommand{\err}{\end{array}}
\newcommand{\bc}{\begin{center}}
\newcommand{\ec}{\end{center}}

\newcommand{\hm}{\,h^{-1}{\rm Mpc}}

\newcommand{\vvec}{{\bf v}}

\newcommand{\rvec}{{\bf r}}

\newcommand{\svec}{{\bf s}}

\def\etal {\rm {\it et al.}}
\def\h0 {\rm H_{0} }

\def\kms{\ifmmode\,{\rm km}\,{\rm s}^{-1}\else km$\,$s$^{-1}$\fi}
\def\hmpc{\ifmmode\,{\it h }^{-1}\,{\rm Mpc }\else $h^{-1}\,$Mpc\,\fi}

\def\\{\hfill\break}

\def\iras{{\it IRAS}}
\def\pscz{{\it PSCz}}
\def\jy{1.2-Jy}
\def\markiii{{\it Mark}\,III}

\def\kms{\ifmmode\,{\rm km}\,{\rm s}^{-1}\else km$\,$s$^{-1}$\fi} 
\def\kms{\,{\rm km\,s${^{-1}}$}}

\def\ltsima{$\; \buildrel < \over \sim \;$}
\def\lsim{\lower.5ex\hbox{\ltsima}}
\def\gtsima{$\; \buildrel > \over \sim \;$}
\def\gsim{\lower.5ex\hbox{\gtsima}}

 \def\v0{\pmb{$0$}}

\def\hm{\ifmmode\,{\it h }^{-1}\,{\rm Mpc }\else $h^{-1}\,$Mpc\,\fi}

\def\baselinestretch{0.96}

\begin{document}

\def\head{
\title{ \iras\, {\it PSCz} {\it v.s.} \iras\, 1.2-Jy Model Velocity Fields: 
A Spherical Harmonics Comparison.}

\lefthead{Teodoro, Branchini \& Frenk}
\righthead{{\it PSCz} {\it v.s.} 1.2-Jy Velocity and Density Fields}

\author {Lu\'{\i}s Teodoro\altaffilmark{1}, Enzo Branchini\altaffilmark{2} and  Carlos S. Frenk\altaffilmark{3}}
\affil{$^1$T-8, Theoretical Division, Los Alamos National Laboratory, Los Alamos, New Mexico, 87545, USA}
\affil{$^2$Dipartimento di Fisica, Universit\`{a} degli Studi  Roma TRE, Roma, Italia}
\affil{$^3$Department of Physics, University of Durham, Science Laboratories, Durham DH1 3LE, England.}
\email{teodoro@lanl.gov,branchin@fis.uniroma3.it and C.S.Frenk@durham.ac.uk}


\begin{abstract}
We have used the two \iras~redshift surveys, \jy~(Fisher \etal~1995)
\nocite{Fisher:1995a} and \pscz~(Saunders \etal~2000)\nocite{Saunders:2000}, 
to model the linear velocity fields within
a redshift of 8\,000 \kms~and have compared them in redshift space.
The two velocity fields only differ significantly in their monopole components.
The monopole discrepancy cannot be solely ascribed to shot-noise errors and
incomplete sky coverage. The mismatch seems to arise from incompleteness
of the \pscz~catalog at fluxes $\le$~1.2 Jy. 
The \jy~and \pscz~higher order velocity multipoles, particularly the 
dipole and quadrupole components,  
appear to be consistent, suggesting that the dipole
residuals found by Davis, Nusser and Willick (1996)\nocite{Davis:1996} 
when comparing \jy~and \markiii~velocity fields probably originates
from the \markiii~velocities calibration procedure rather than from
uncertainties in the model velocity field.
Our results  illustrate the efficiency of the spherical harmonics decomposition techniques
in detecting possible differences between real and model velocity fields.
Similar analyses shall prove to be very useful in the future to test the reliability of 
next generation model velocity fields derived from new redshift catalogs
like 2dFGRS (Colless \etal~2001\nocite{Colless:2001}), 
SDSS (York \etal~2000)\nocite{York:2000}, 6dF and 2MRS.
\end{abstract}

\keywords{
cosmology ---
theory -- galaxies
large-scale structure of the Universe --- 
large-scale dynamics.
}
}
\twocolumn[\head]


\section{Introduction}\label{ch:comparison}

In the framework of linear gravitational instability theory and
linear biasing a comparison between the predicted and observed peculiar
velocities allows us to measure the so called $\beta$ parameter
\begin{equation}
\beta \equiv \frac{\Omega_m^{0.6}}{b}.
\end{equation}
i.e. a combination of the mass density parameter, $ \Omega_m$,
and the linear bias parameter, $b$, which relates the mass overdensity
field $\delta$ to the density fluctuations in galaxy counts $\delta_g$,
through $\delta_g=b\delta$. 
Several different techniques have been used to compare measured peculiar velocities
(taken from all available catalogs: \markiii~Willick 1997a\nocite{Willick:1997a}, 
SFI Haynes 1999\nocite{Haynes:1999}, ENEAR da Costa \etal~2000\nocite{daCosta:2000},
SEcat Zaroubi 2002\nocite{Zaroubi:2002})
to predict velocities that, in most cases, have been obtained from
two redshift surveys of IRAS galaxies: the \jy~and the \pscz~catalogs.

Most analyses have investigated the consistency between model 
and observed velocities, while only a few of them have been devoted
in comparing different datasets or in studying the consistency of 
model velocity fields.
All recent works show a general consistency between models and data
(Willick \etal\, 1997, hereafter WSDK\nocite{Willick:1997}; 
Willick and Strauss 1998\nocite{Willick:1998}; 
Sigad \etal~1998\nocite{Sigad:1998};
Dekel \etal~1999\nocite{Dekel:1999};
Branchini \etal~1999, hereafter B99\nocite{Branchini:1999}; 
Nusser \etal~2000\nocite{Nusser:2000};
Zaroubi \etal~2002\nocite{Zaroubi:2002})
with one noticeable exception represented by the work of
Davis, Nusser \& Willick (1996, hereafter DNW)\nocite{Davis:1996}.
Their analysis showed that the velocity residuals between 
the \markiii~peculiar velocities and the \jy~predictions 
display a significant spatial correlation.
Subsequent analyses by WSDK and Willick and Strauss (1998)\nocite{Willick:1998}
were performed using the same dataset, a similar model velocity field but 
a different comparison technique, called VELMOD, which allows for 
independent calibrations of \markiii~velocities
in each of the \markiii~sub-catalogs.
As a result the observed 
and model velocity fields were brought into better agreement, at least within
a redshift distance $s = 7\,500$\kms~and returned a
calibration inconsistent with the original one performed
by Willick \etal~(1997a)\nocite{Willick:1997a}.

While these results suggest that the DNW mismatch originates
from calibration problems in the \markiii~catalog, the possibility
that systematic errors in the \jy~model velocities also 
contribute to the discrepancy has been completely overlooked.
In particular, we ignored whether the velocity fields predicted from
the deeper  IRAS \pscz~catalog provides a better match to
the \markiii~velocities than the shallower \jy~catalog, 
considered by DNW.

The main goal of this work is to check whether the discrepancies between
the \jy~and \markiii~velocities can be ascribed, at least to some extent,
to model uncertainties. For this purpose we compare both the 
overdensity and the radial  velocity fields modeled from
the redshift space distribution of \jy~and  \pscz~galaxies.
This work is meant to stress the 
usefulness of spherical harmonics 
decomposition techniques in assessing the adequacy of a model
velocity (or density) field through 
velocity-velocity (or density-density) comparisons.

In \S~\ref{section:decomposition5} we describe how to compare
velocity and overdensity fields inferred from different redshift 
surveys. The galaxy redshift catalogs used in our analyses are described in
\S~\ref{section:datasets}. In sections \ref{section:fields5} and 
 \ref{section:radialvelocitymaps} we analyse 
the two model overdensity fields and radial velocity fields.
Our main results are discussed in \S~\ref{section:discussion5} 
and our main conclusions are presented in
\S~\ref{section:conclusions5}. 

\section{Spherical Harmonic Coefficients Decomposition} \label{section:decomposition5} 

The relations between an arbitrary scalar field  $\psi(\svec)$ 
and the real--valued spherical harmonic coefficients $\psi_{lm}(s)$ 
are
\begin{equation}
\psi_{lm}(s) = \int \psi(\svec){\mathcal Y}_{lm}(\hat{\svec})d\Omega
\label{eq:sphcoeff5}
\end{equation}
\begin{equation}
\psi(\svec)=\sum_{l=0}^{\infty}\sum_{m=-l}^{l}\psi_{lm}(s){\mathcal Y}_{lm}({\hat \svec}),
\label{eq:sphericalhar5}
\end{equation}
where  ${\mathcal Y}_{lm}(\hat{\svec})\{l=0,..,\infty; |m|\le l \}$
denote the well known real-valued spherical harmonics, defined as in 
Baker \etal~(1999, hereafter BDSLS)\nocite{Baker:1999} and 
Bunn (1995)\nocite{Bunn:1995} (see also Jackson 1999\nocite{Jackson:1999}).
For a given $l$~these  functions are normalized to the value of
${\mathcal{Y}}_{l0}$ at the North Galactic Pole.

The amplitude of the $l$-multipole $\psi_{l}(s)$ is 
defined by the sum in quadrature over the $(2l+1)$-$\psi_{\,lm}$'s, where $\{m=-l,...,l\}$:
\begin{equation}
\psi_l(s) = \left [ \sum _{m=-l}^{m=l} \psi^2 _{lm} (s) \right ] ^{\frac{1}{2}}.
\end{equation}

\subsection{Peculiar velocity and overdensity fields from the
  distribution of galaxies in redshift space}\label{subsec:pec_vel_dens_gal_red_spac}

Nusser \& Davis (1994)\nocite{Nusser:1994} show that in the linear regime 
the peculiar velocity field is irrotational not only
in real space but also in redshift space
and thus can be expressed as the gradient of a scalar velocity potential:
$\vvec(\svec)=-\nabla \Phi(\svec)$.  
At a given redshift $s$,
$\Phi(s)$ and $\hat{\delta}(s)$ can be expanded in spherical harmonics
and related to each other through a modified Poisson equation:
\begin{equation}
{{1}\over{s^2}}
\left( s^2 {\Phi_{lm}'} \right)'
-{{1}\over{1+\beta}} {{l(l+1)\Phi_{lm}}\over{s^2}}=
{{\beta}\over{1+\beta}}
\left( \hat{\delta}_{lm}-{{1}\over{s}}{{d\ln{\phi}}\over{d \ln{s}}}
{{ \Phi_{lm}}' } \right),
\label{eq:ndchapter5}
\end{equation}
where $\phi(s)$ and $\hat{\delta}_{lm}({\svec})$ 
denote the sample selection function and the spherical components 
of the overdensity field, respectively. Here prime expresses $d/ds$.
Solving this differential equation requires computing the redshift space density field 
from the observed galaxy distribution on a spherical grid using Gaussian 
cells of approximately equal solid angle. 
These cells are equally distributed in longitude (64 bins) and are centered
at the Gaussian-Legendre 32 point quadrature formula in the range 
$ -1 \le \cos(b) \le 1$, where $b$ is the Galactic latitude.
In this work we measure redshifts in the Local Group frame, and use 52 Gaussian 
radial redshift bins out to a distance of $s=18\,000$\kms.
When comparing two or more galaxy catalogs
the size of the radial bin is set by requiring 
equal spatial resolution and the minimal shot noise constant 
throughout the volume.
We do this setting the distance between the centers of the 
Gaussian cells at distance $s$ equal to the average galaxy-galaxy separation, 
$\sigma(s)=[\bar{n}\phi(s)]^{-1/3}$, in the sparser catalog.
The mean number density is estimated as 
\begin{equation}
\bar{n}=\frac{1}{V} \sum _i\frac{1}{\phi(s_i)},
\label{eq:meandensity5}
\end{equation}
where the sum is over the galaxies contained within
the volume $V$, which, in this work, is a sphere of
radius $8\,000$\kms.

\begin{figure}[t]
\begin{center}
\scalebox{0.43}{\includegraphics[38,172][587,693]{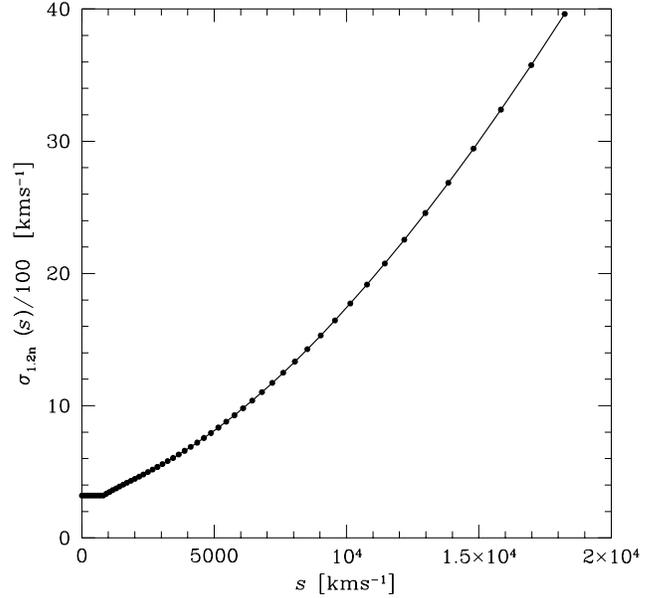}}
\caption{ The \jy~smoothing width as a  function of redshift. The dots
represent the value of $\sigma_n$ at 
centre of the 52 redshift bins.\label{fig:fig1}}
\end{center}
\end{figure}

Hernquist and Katz (1989)\nocite{Hernquist:1989} pointed out that there are
two possible ways to estimate the smoothed density field: 
the "scatter" and "gather" approches. In the former approch each particle 
is distributed in space, and the density estimate 
at a given point is the superposition of individual smoothing spheres.
In the latter case one defines a smoothing lenght at a given point and weights
the particle in its neighborhood by the resulting kernel.
The Gaussian-smoothed galaxy overdensity
field at a grid cell centred around ${\svec}_n$  is given by:
\begin{equation}
1+{\hat \delta}(\svec_n)=\frac{1}{(2\pi)^{3/2}\sigma_{n}^3}
\sum_i^{N} \frac{1}{\phi(s_i)}\exp\left [
  -\frac{(\svec_n-\svec_i)^2}{2\sigma_{n}^2} \right ],
\label{eq:smoothing5}
\end{equation}
where the sum is over the $N$ galaxies in the catalog, which clearly is 
a "gather" approach.
The smoothing width at a redshift $s$, used in this paper is
$\sigma_{n}=\max[ 320, \sigma(s)]$ \kms.
Since in this work we compare the \pscz~catalog
with the sparser  \jy~catalog, we use the 
$\sigma_n$ obtained from the \jy~catalog shown in Fig.~(\ref{fig:fig1}).
This smoothing scheme is tailored to
keep the shot-noise uncertainty roughly constant throughout the sample volume
and is similar to the optimal Wiener filtering procedure used by Lahav
\etal~(1994)\nocite{Lahav:1994} and  Fisher \etal~(1995b)\nocite{Fisher:1995b}. 
Moreover, the our kernel has a compact support, which simplifies the 
computational task.

Eqn.~(\ref{eq:ndchapter5}) is valid in the linear regime, where  
a one-to-one mapping between distance and redshift is guaranteed.  
This assumption is not valid in high density regions, 
such as cluster of galaxies, where shell crossing may occur 
triple valued-regions may apear.
To deal with the latter effect, we adopt the same
procedure as Yahil \etal~(1991)\nocite{Yahil:1991} and collapse the
fingers of God associated with the six richest clusters in the sample listed in
Table 2 of that work.

In this work we will consider galaxies within 
$s_{\mbox{\scriptsize{max}}} = 20\,000$ \kms~
with the spherical harmonics decomposition limited to 
$l_{\mbox{\scriptsize{max}}}=15$.
Outlying $s_{\mbox{\scriptsize{max}}}$ galaxies are assumed to be 
distributed uniformly according to the mean number density of the 
parent catalogs.

\section{Datasets}
\label{section:datasets}

The  \iras~\pscz~catalog
contains $\approx$ 15\,500 \iras~PSC galaxies with a 60
$\mu$m flux larger than 0.6 Jy. The average depth of this survey is $\approx$ 100
\hm.  We restrict  our analysis to a sub-sample of 11\,206 galaxies 
within 20\,000 \kms~from the Local Group (LG, hereafter). 
The areas not covered by the survey amount to 16.0\% of the sky and are
preferentially located near the galactic plane, in the 
so-called Zone of Avoidance (Saunders \etal~2000\nocite{Saunders:2000}).
The empty areas have been filled-in with the cloning 
procedure described in Branchini \etal~(1999) 
\nocite{Branchini:1999}.

The \jy~catalog (Fisher \etal~1995a\nocite{Fisher:1995a}) contains 5\,331 \iras\ 
galaxies with a 60 $\mu$m flux larger than 1.2-Jy within 20\,000 \kms.
This catalog has a slightly larger sky coverage of ($\approx$\ 87.6$\%$) and a smaller 
median redshift ($\approx$ 84 $\hm$).
The same B99 filling technique has been used to restore
all sky coverage.

Fisher \etal~(1995b)\nocite{Fisher:1995b}
have found that uncertainties in filling-in the
empty areas of the \jy~survey
do not appreciably affect the spherical harmonics coefficients
with $ l \la 15$. Similar conclusions are also valid
for the \iras~\pscz~survey (Teodoro \etal~1999)\nocite{Teodoro:1999}.

The selection functions of the two surveys, that quantify the probability
of a galaxy at distance $s$, have been determined as  maximum-likelihood  
fits to the smooth function
\begin{equation}
\phi(s) = \left \{ \begin{array}{ll} \left (\frac{r_r}{r}\right)^{2\alpha} \left(\frac{r_{\star}^2+r^2_s}{ r_{\star}^2+r^2} \right)^\beta &\quad,~ s \ge H_0\,r_s, \\
 					1& \mbox{\quad,~ otherwise,}\\
\end{array}
\right .
\label{eq:selfunct}
\end{equation}
where $r = s/H_0$\footnote{Throughout this article we write Hubble's 
constant as $H_0=100${\it h} \kms~Mpc$^{-1}$}~is expressed 
in $\hm$~(Yahil \etal~1991\nocite{Yahil:1991}). 
The best-fit parameters, determined using the galaxies within $8\,000$\kms~and 
assuming a flat universe with $\Omega_m = 1$, 
are shown in Tab.~(\ref{selfpsczsub}). The quantity $r_s$ in 
Eqn.~(\ref{eq:selfunct})
accounts for the incompleteness of faint galaxies in the innermost
regions of the catalogs (Rowan-Robinson \etal~1990,
Yahil \etal~1991)\nocite{Rowan-Robinson:1990,Yahil:1991}.

\begin{table*}[t]
\begin{center}
\caption[]{Selection function parameters for \iras~catalogs.}
\tabcolsep 2pt
\begin{tabular}{ccccccccccc}  \\ \hline \hline
\\
 & Catalog & $f_{\mbox{\scriptsize{$60\mu m$}}}^{\mbox{\scriptsize{lim}}}$ &  N${gal}^{\mbox{\,{\scriptsize{a}}}}$ & ~~~~$\alpha$~~ &~~$\beta$~~ &~~$\gamma$~~& ~~$r_s^{\mbox{\,{\scriptsize{b}}}}$~~&~~$r_{\star}$~~& ~~~~$\bar n $~~~~~ & \\  
 &  & [Jy] & & & & & [$\hm$] & [$\hm$] & $[h^3\mbox{Mpc}^{-3}]$& \\
\\
\hline
\\
&\pscz& 0.6 & 13\,364 & 0.54 & 1.80 & ...& 6.0 & 87.00~ & ~6.12 $\cdot10^{-2}$ &\\ 
&\pscz$_{~0.7}$ & 0.7 & 11\,170 & 0.57 & 1.80 & ... & 6.0 & 76.35~ & ~6.38 $\cdot10^{-2}$ &\\ 
&\pscz$_{~1.2}$ & 1.2 & 5\,519 & 0.43 & 1.86 & ... & 5.0 & 50.40~ & ~4.62 $\cdot10^{-2}$ & \\ 
&1.2-Jy & 1.2 & 6\,010 & 0.43 & 1.86 & ...& 5.0 & 50.60~ & ~4.67 $\cdot10^{-2}$ & \\
&\pscz$_{\mbox{\scriptsize{\,evo.}}}$ $^{\mbox{\hspace{-6mm}\,{\scriptsize{c}}}}$ & 0.6 & 13\,364 & 0.99 & 3.45 & 1.93 & 6.0  & 76.02~ &  ~5.76 $\cdot10^{-2}$ &\\ 
\\
\hline \hline
\end{tabular}
\label{selfpsczsub}
\end{center}
\flushleft
\small{
\hspace{38.0mm}{\scriptsize{$^{\mbox{\,{\scriptsize{a}}}}$}}\ Includes synthetic objects.\\
\hspace{38.0mm}{\scriptsize{$^{\mbox{\,{\scriptsize{b}}}}$}}\ Parameter kept constant in the likelihood maximization.\\
\hspace{38.0mm}{\scriptsize{$^{\mbox{\,{\scriptsize{c}}}}$}}\ The selection function parameters have been taken from Canavezes \etal\ (1999) \nocite{Canavezes:1998}}\\
\end{table*}

\begin{table*}[t]
\begin{center}
\caption{The parameters of the $\delta_{1.2}-\delta_{0.6}$\ and $u_{1.2}-u_{0.6}$\ linear regressions.}
\tabcolsep 1pt
\begin{tabular}{cccccc} \\\hline \hline
\\
~~Regression~~ & {{$N_{\mbox{\scriptsize{grid}}}$}~~~}&
{~~~{$N_{\mbox{\scriptsize{dof}}}$}~~~}  &{{~~~~~~~~~~$A^{\mbox{\,{\scriptsize{a}}}}$~~~~~~~~~~}} &
{~~~~~~$B$~~~~~}&{ ~~~$\chi^2_{\mbox{\scriptsize{eff}}}/N_{\mbox{\scriptsize{dof}}}$~~~} \\  
\\
\hline
\\
$\delta_{1.2}-\delta_{0.6}$ & 112\,552 & 156.36  & ~~$-0.011\pm 0.014$~~   &  ~~$1.025\pm0.036$~~ & 0.78 \\
$u_{1.2}- u_{0.6}$ & 112\,552 & ~$\la$ 156.36~  & $60.0\pm 5.9$ & $1.040\pm 0.019$ & 0.88 \\
\\
\hline \hline
\end{tabular}
\\
\label{t:comp}
\end{center}
\flushleft
\small{
\hspace{39mm}Notes:\\
\hspace{39mm}{\scriptsize{$^a$}} For the $u-u$ comparison $A$ is expressed in \kms.\\
\hspace{39mm}Column 2: $N_{\mbox{\scriptsize{grid}}}$, the number of gridpoints used for the regression;\\
\hspace{39mm}column 3: $N_{\mbox{\scriptsize{dof}}}$, the number of independent volumes;\\
\hspace{39mm}column 4: $A$, the zero--point offset in the linear regression, and its $1-\sigma$\ error;\\
\hspace{39mm}column 5: $B$, the slope of the linear regression, and its $1-\sigma$\ error;\\
\hspace{39mm}column 6: $\chi^2_{\mbox{\scriptsize{eff}}}/N_{\mbox{\scriptsize{dof}}}$ from the linear regression.}
\end{table*}

\section{The \iras~\pscz~and \jy~Fields Overdensity}
\label{section:fields5}

The first step of our analysis consists of applying
Eqn.~(\ref{eq:smoothing5}) to compute the
\iras~\jy~and \pscz~overdensities onto a spherical grid of radius 
$18\,000$ \kms~and $l_{\mbox{\scriptsize{max}}}=15$ with
the smoothing scheme given by Eqn.~(\ref{eq:smoothing5}).

\subsection{Overdensity Maps}\label{subsec:overdensity_maps}

\begin{figure}[t]
\begin{center}
\scalebox{0.40}{\includegraphics[31,313][569,580]{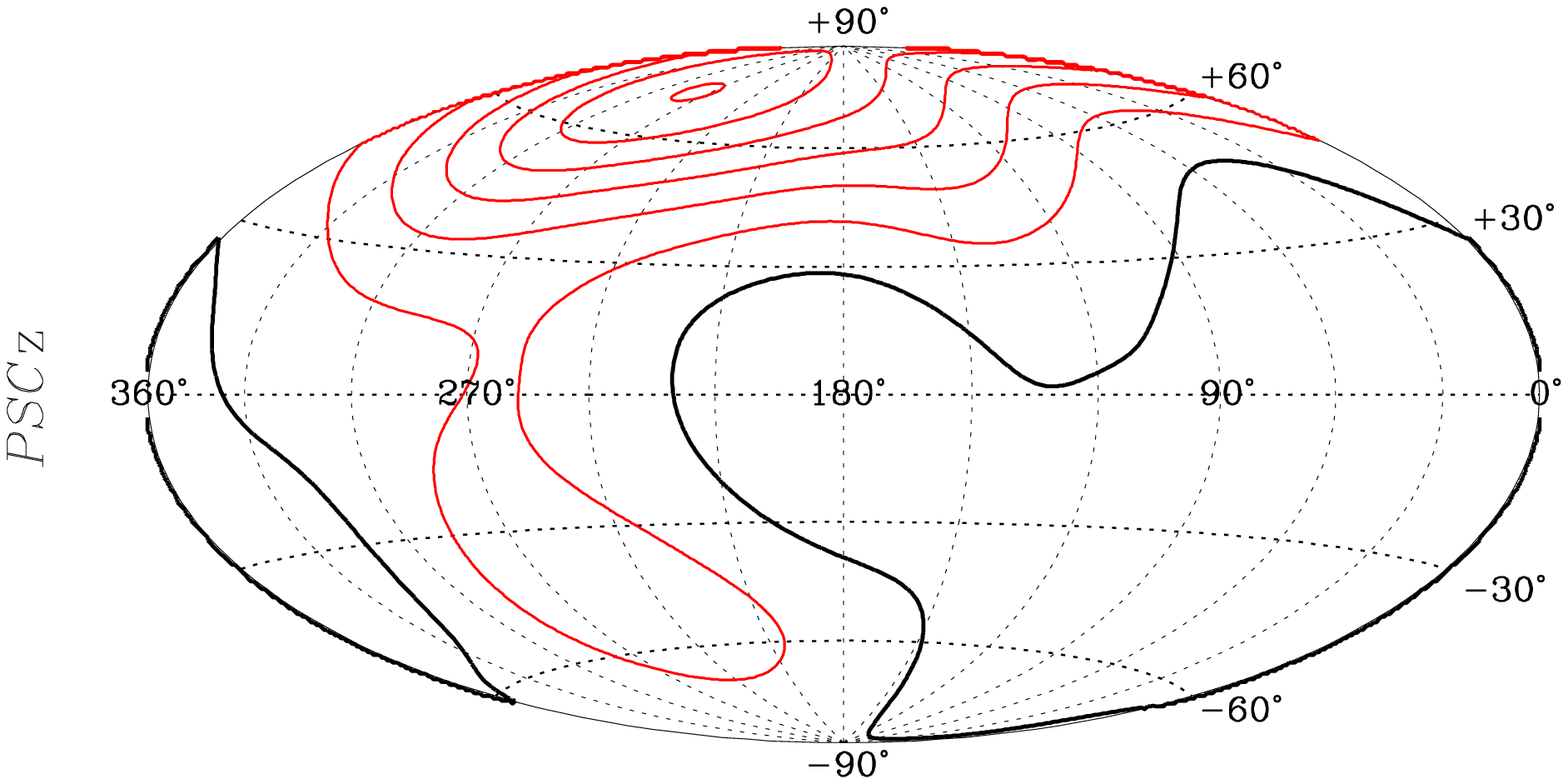}}
\scalebox{0.40}{\includegraphics[31,313][569,580]{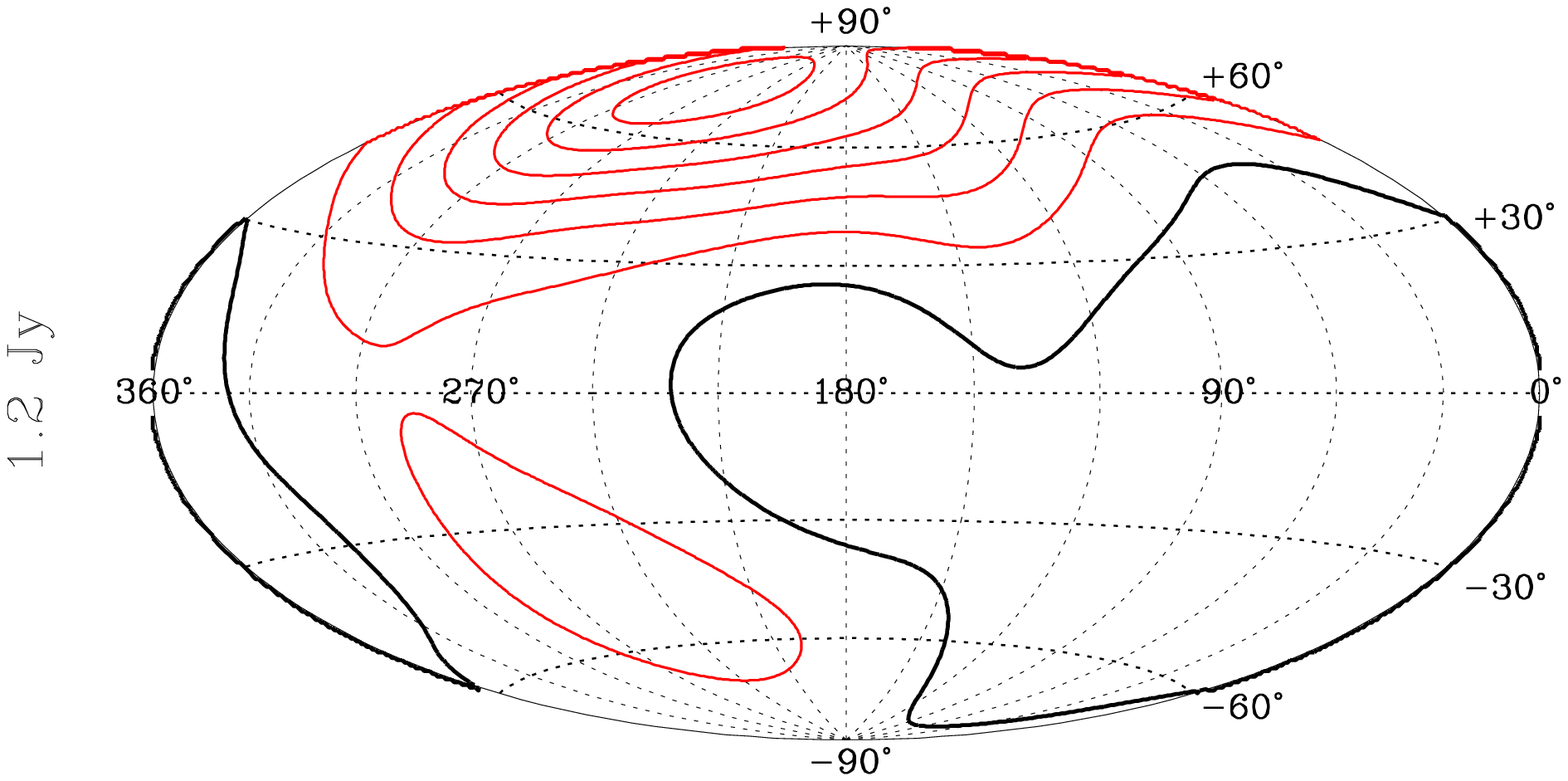}}
\scalebox{0.40}{\includegraphics[31,263][569,580]{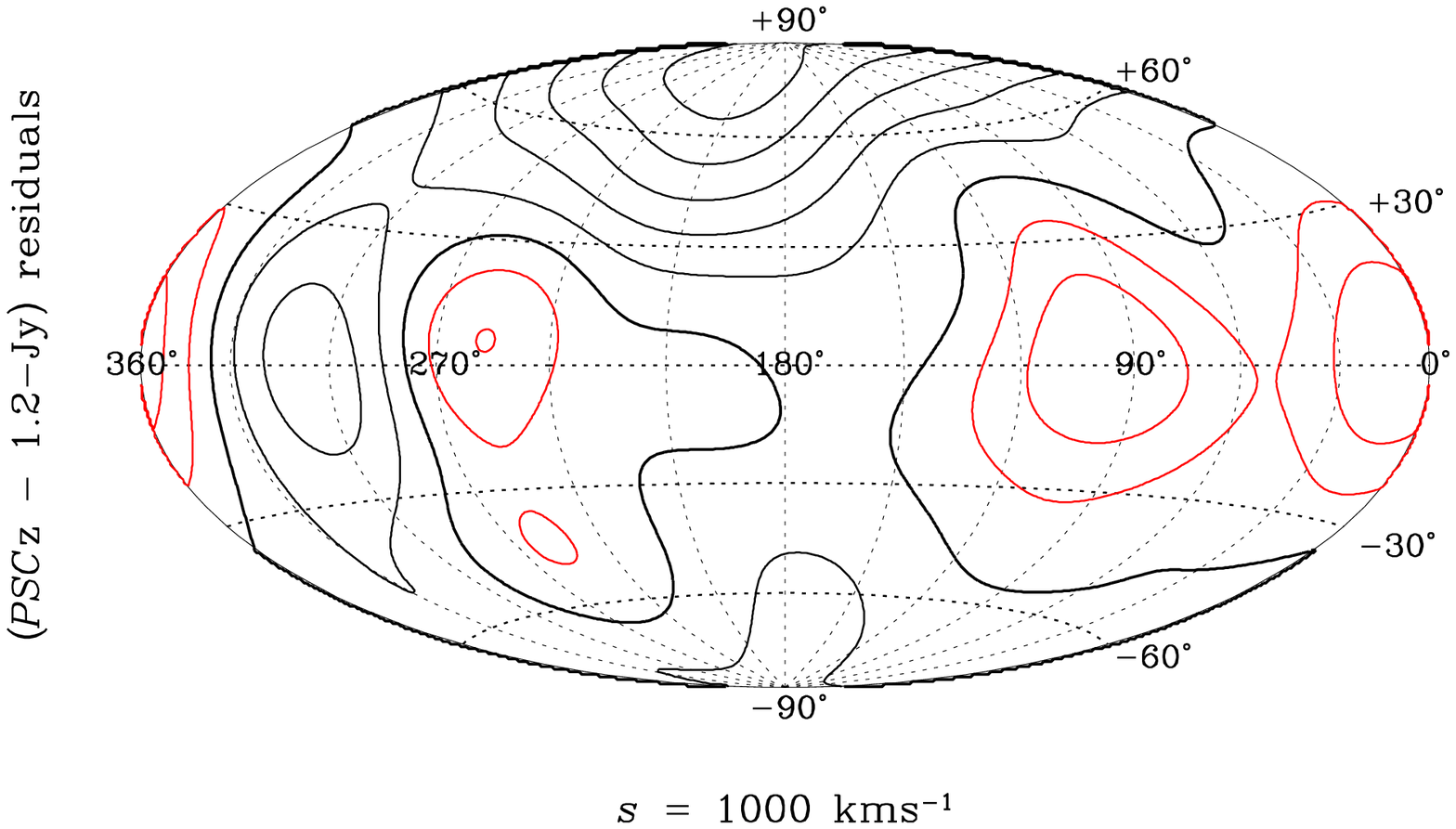}}
\caption{\pscz~(top), \jy~(middle) and residual (bottom)
overdensity fields in the radial shell centred at $s = 1\,000$\ \kms. 
The thick-line contour indicates $\delta = 0$\ or $\delta_{\mbox{\scriptsize{res}}} = 0$ (bottom).
The light- and dark-gray contours show over- and under-dense regions, respectively. In the top and middle panels
the contours are equally--spaced by $\Delta \delta$\ = 1.0. 
The bottom panel the density residuals are shown in steps of 
$\Delta \delta_{\mbox{\scriptsize{res}}} = 0.125 $. \label{fig:dens1000}}
\end{center}
\end{figure}

\begin{figure}[b]
\begin{center}
\scalebox{0.40}{\includegraphics[31,313][569,580]{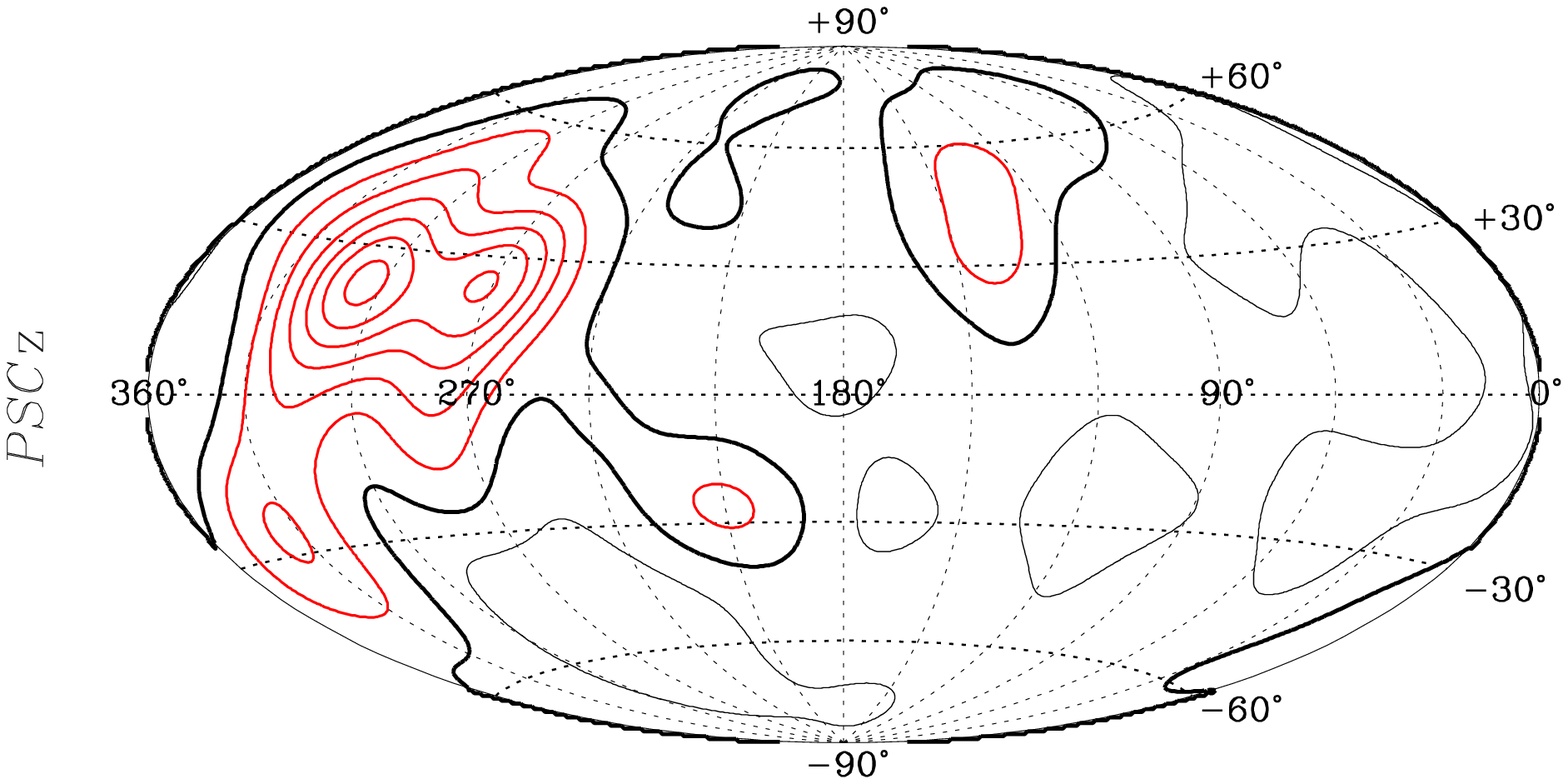}}
\scalebox{0.40}{\includegraphics[31,313][569,580]{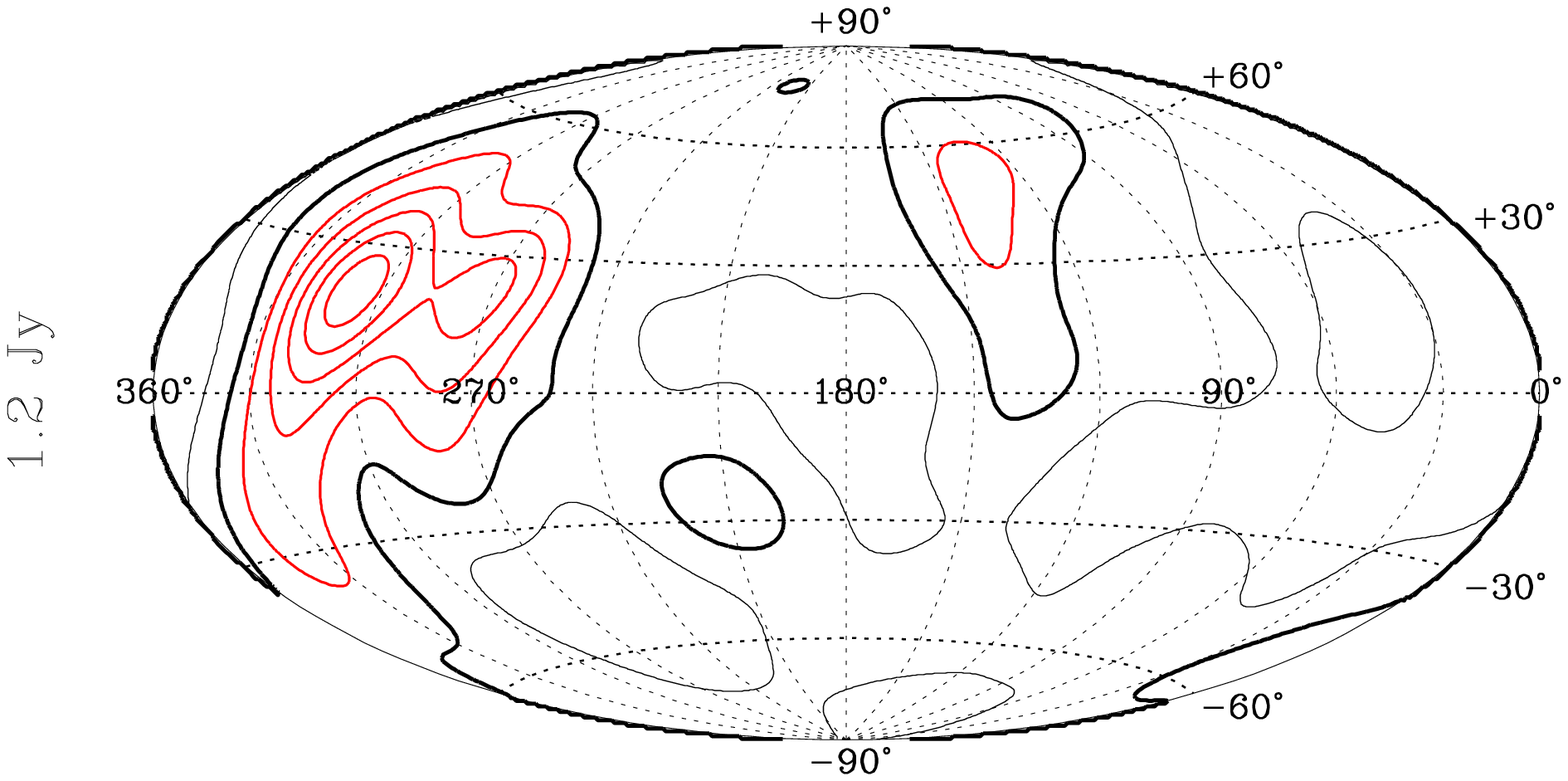}}
\scalebox{0.40}{\includegraphics[31,263][569,580]{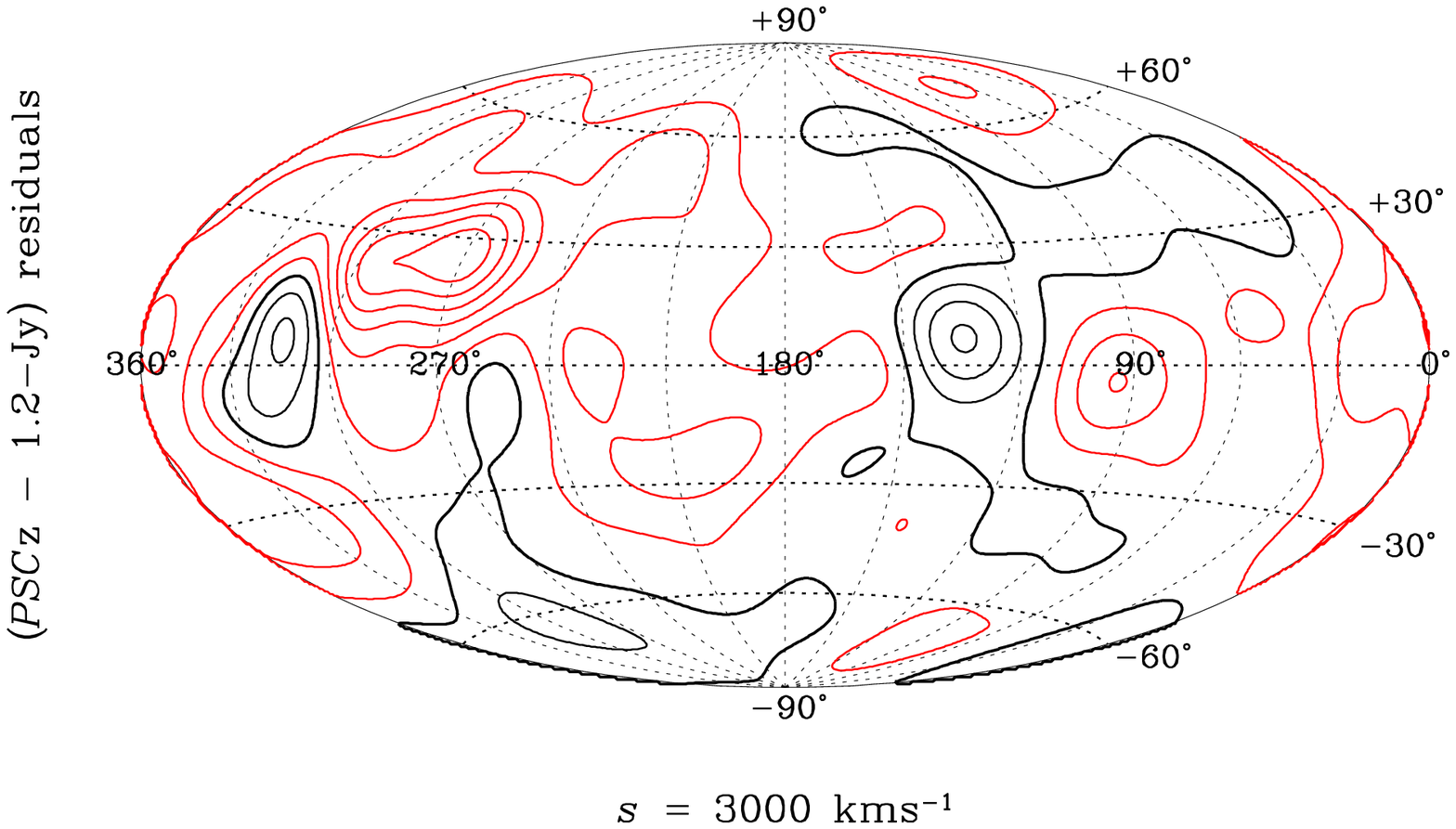}}
\caption{Same as in Fig.~(\ref{fig:dens1000}). The shell is now centred 
at $s = 3\,000$\ \kms~and the contour spacings are $\Delta \delta$\ = 0.5
and $\Delta \delta_{\mbox{\scriptsize{res}}} = 0.125 $.}
\label{fig:dens3000}
\end{center}
\end{figure}

\begin{figure}[b]
\begin{center}
\scalebox{0.40}{\includegraphics[31,313][569,580]{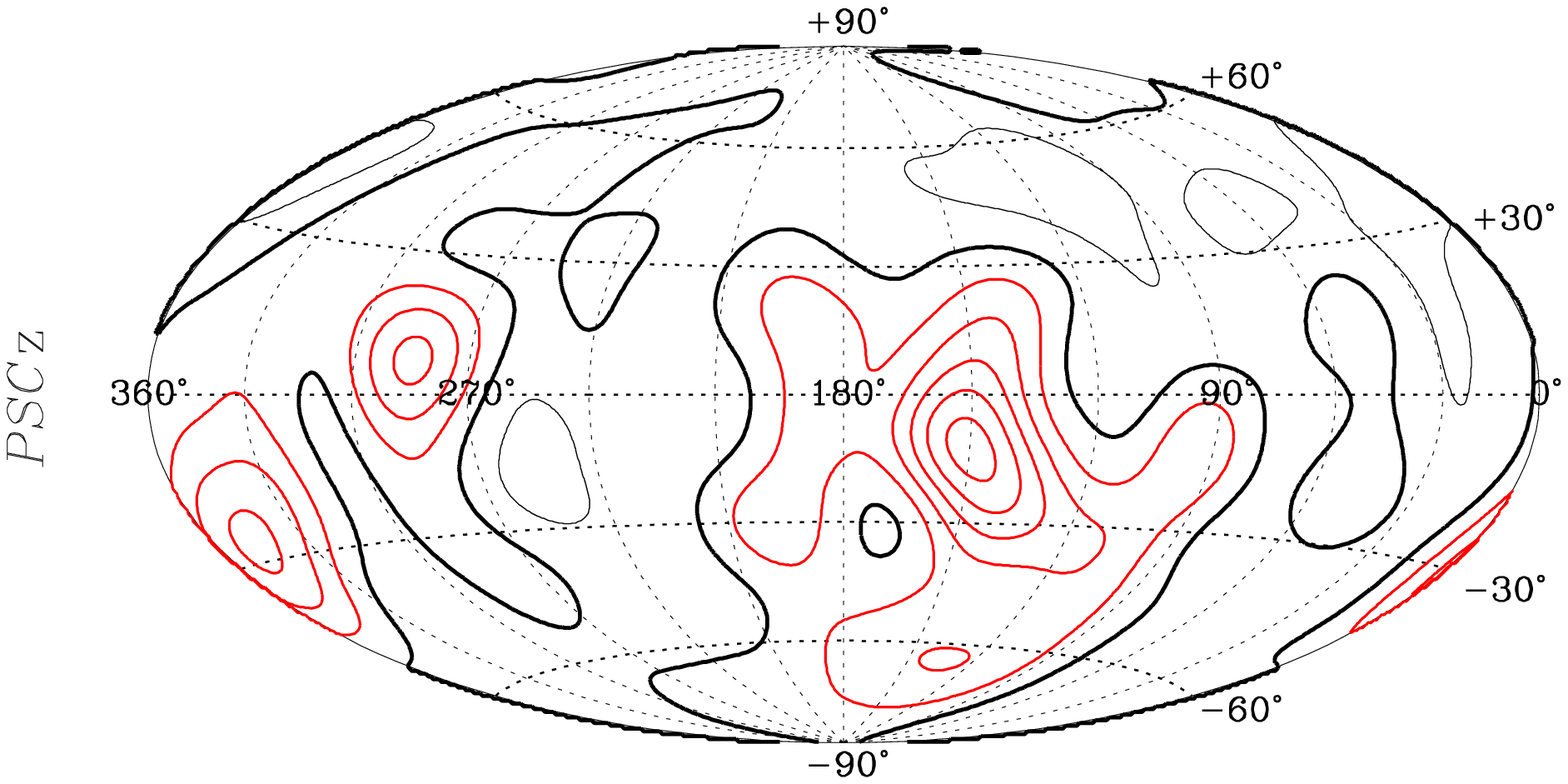}}
\scalebox{0.40}{\includegraphics[31,313][569,580]{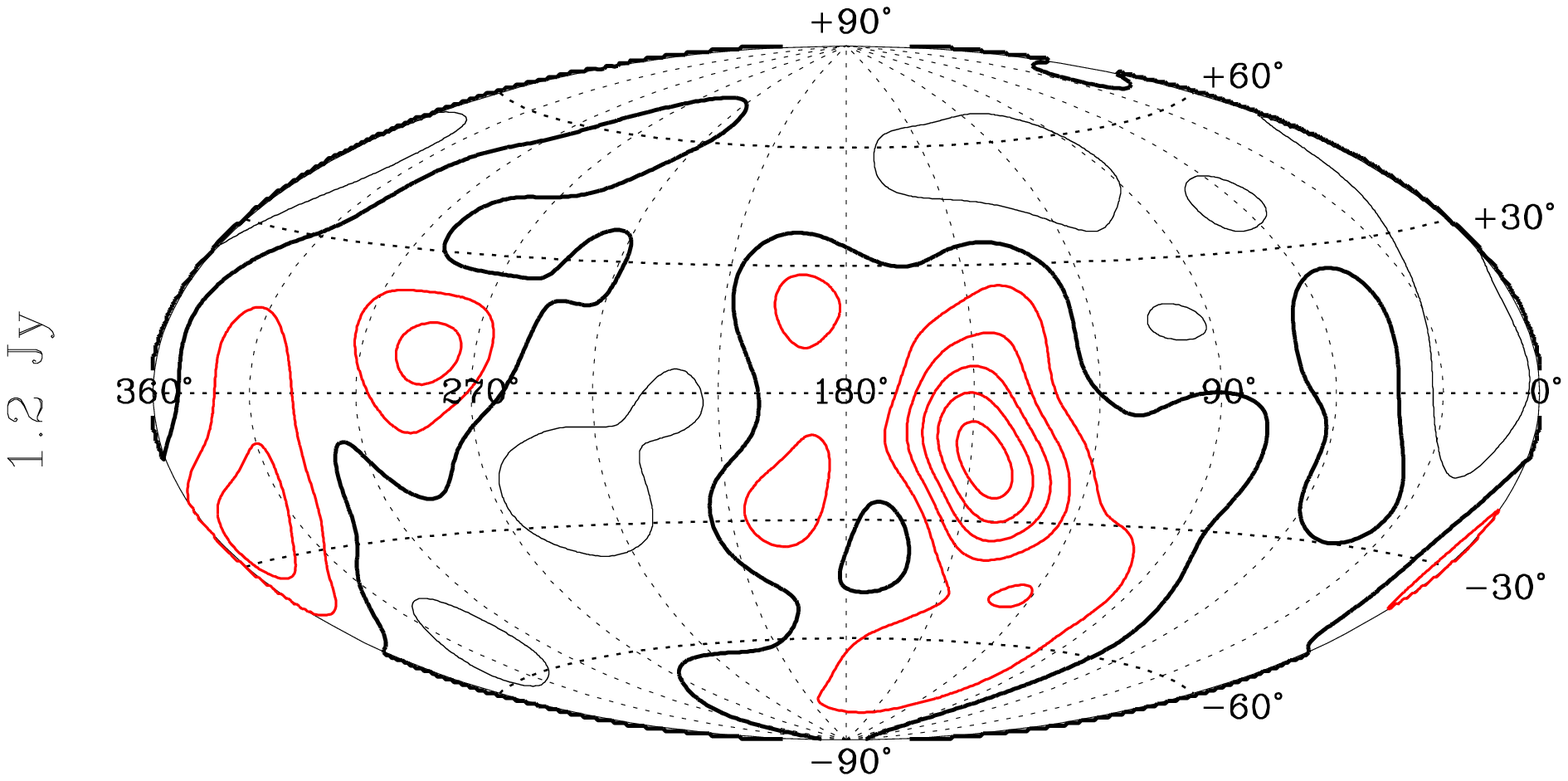}}
\scalebox{0.40}{\includegraphics[31,263][569,580]{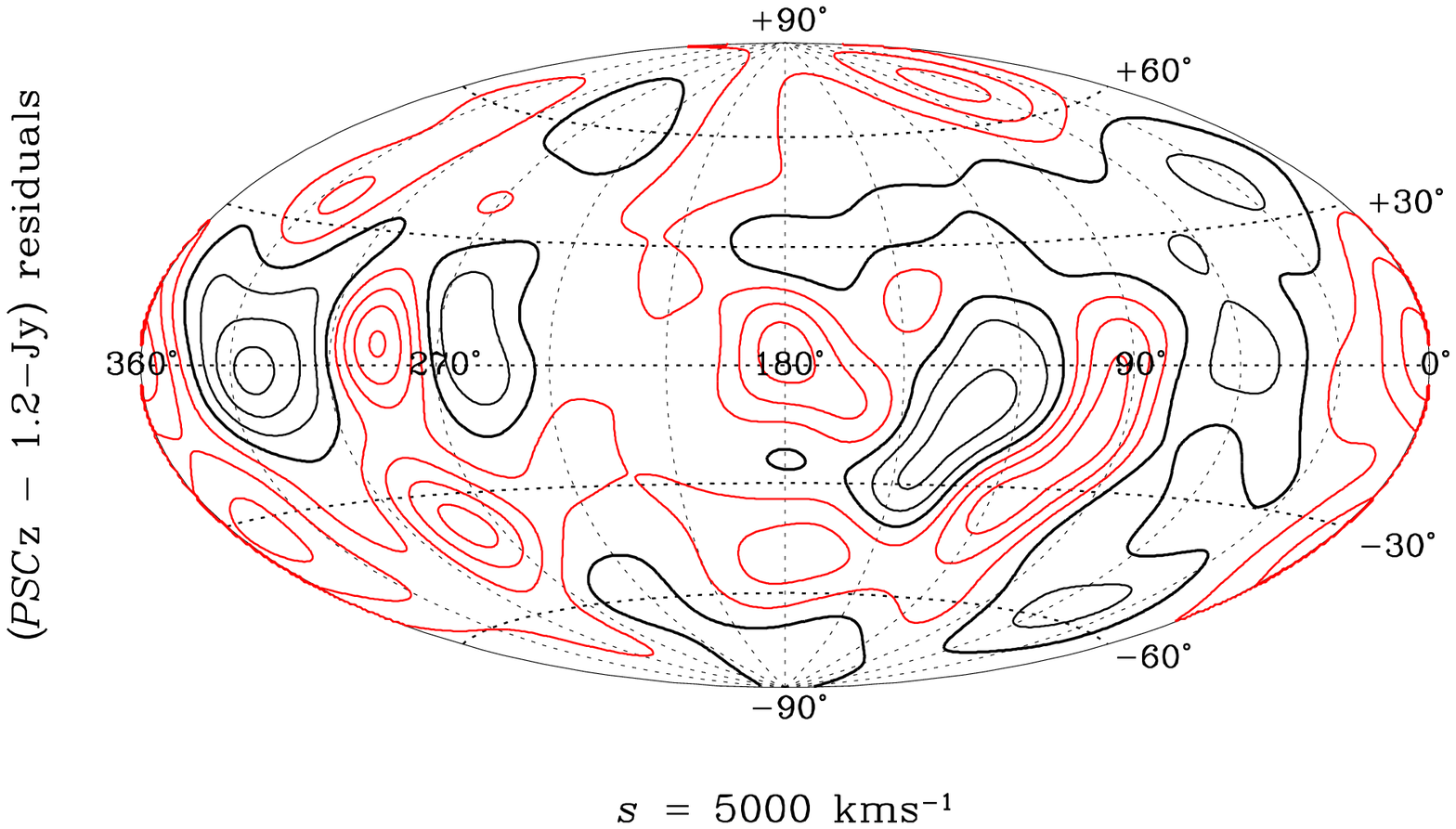}}
\caption{Same as in Fig.~(\ref{fig:dens1000}). The shell is now centred 
at $s = 5\,000$\ \kms~and the contour spacings are $\Delta \delta$\ = 0.5
and $\Delta \delta_{\mbox{\scriptsize{res}}} = 0.125 $.}
\label{fig:dens5000}
\end{center}
\end{figure}

In Figs. (\ref{fig:dens1000}--\ref{fig:dens5000}) 
we show the Aitoff projections of the \pscz~(top panels) 
and \jy~(middle panels) 
galaxy overdensity fields in three spherical shells centred at 
$s=$~1\,000, 3\,000 and 5\,000\kms, respectively.
The thickness of each shell is defined by the Gaussian width 
at the distance of the shell. All maps are plotted 
in Galactic coordinates.
The bottom panels show the map of the overdensity residuals
$\delta_{\mbox{\scriptsize{res}}}(s)=\delta_{0.6}(s)-\delta_{1.2}(s)$,
where the subscripts $_{0.6}$ and $_{1.2}$ refer to \pscz~and
\jy~surveys, respectively.
In all plots the thick line indicates the zero overdensity (or residual) contour.
The overdensity (residual) contour spacing $\Delta \delta$ 
($\Delta \delta_{\mbox{\scriptsize{res}}}$) depends on the shell's depth and 
is indicated in the Figure captions.
Detailed cosmographic studies of the structures in the nearby universe 
have already been performed using both the distribution of \jy~(Strauss 
and Willick 1995\nocite{Strauss:1995a}) and \pscz~galaxies\footnote{See 
\url{http://www-astro.physics.ox.ac.uk/$\sim$wjs/pscz.html}} 
and will not be repeated here. We only stress that the main cosmic structures
like the Virgo Cluster 
$(l \approx   240^{\circ},\ b \approx  75^{\circ}, \ s=1000 $\kms$)$,
the local void spanning the region
$ 330 ^{\circ} \la l \la 120 ^{\circ},\ -90^{\circ} \la b \la 30 ^{\circ}$
at $ s=1000$ \kms, the Hydra-Centaurus complex 
($l \approx 270^{\circ},$ $\ b\approx 20^{\circ},\ s=3000 $\kms$)$
and ($l\approx 300^{\circ},\ b\approx 20^{\circ},\ s=3000$ \kms $)$
and the Perseus-Pisces supercluster 
($l \approx  140^{\circ},\ b\approx -25^{\circ},\ s=3000 $ \kms $)$
are detected in both surveys.

The residuals, calculated at the same gridpoints as the density fields, 
seem to indicate that, as expected, the largest residuals are located 
near the galactic plane
where uncertainties in the filling-in procedure dominate the errors.
In the inner shell, overdensities in the \jy~galaxy distribution appear to 
be larger than the \pscz~ones while in the most external shell the
situation is reversed.

\subsection{Density Residuals}\label{section:dresiduals}

A way of quantifying possible systematic differences 
between the two fields is that of performing a point-to-point
comparison. In absence of systematic errors we expect the 
two IRAS overdensity fields to be linearly related through
\begin{equation}
\delta _{1.2} = B \delta _{0.6} + A
\end{equation}  
with $A=0$ and $B=1$. $\delta _{0.6}$ and $\delta _{1.2}$ are
estimated at Cartesian gridpoint positions.
A zero--point offset $A \neq 0$ indicates  discrepancies in the mean densities
of the two catalogs while a slope $B \neq 1$ reveals the presence of 
systematic errors.
If the values of $\delta_{0.6}$\ and $\delta_{1.2}$\ are independent and 
their errors  are Gaussian then we can compute the $\chi^2$ statistics
\begin{equation}
\chi^2=\sum_{i=1}^{N_{\mbox{\scriptsize{grid}}}} 
\frac 	{ \left (\delta_{1.2}-A-B\delta_{0.6}\right )^2 } 
	{ \left (\sigma_{1.2}^2 + B^2 \sigma_{0.6}^2 \right )},
\end{equation}
where $\sigma_{0.6}$\ and $\sigma_{1.2}$\ indicate the typical 
errors in $\delta$ and ${N_{\mbox{\scriptsize{grid}}}}$~is 
the total number of points in the comparison. However, since 
the gridspacing (250\kms)
is smaller than the smoothing length ($\ge 320$\kms), not all values 
of $\delta_{0.6}$ (or $\delta_{1.2}$) are independent. Indeed, the 
effective number of independent  points used in the
$\chi^2$ statistics, $N_{\mbox{\scriptsize{eff}}}$ is given by
\begin{equation}
N_{\mbox{\scriptsize{eff}}}^{-1} = {N_{\mbox{\scriptsize{grid}}}}^{-2}
\sum _{i=1}^{N_{\mbox{\scriptsize{grid}}}}
\sum _{j=1}^{N_{\mbox{\scriptsize{grid}}}}
\exp (-r_{ij}^2/2R^2_{s,i}),
\label{eq:neff}
\end{equation}
where $r_{ij}$ is the separation between gridpoints $i$ and $j$ and $R_{s,i}$\ is the 
smoothing radius of the Gaussian kernel at the gridpoint $i$. This expression is 
a generalization of Eqn.~(12) in Dekel \etal~(1993)\nocite{Dekel:1993} to 
the case of a grid with variable mesh-size since $R_{s,i}$ depends on 
the gridpoint position.
The new statistics $\chi^2_{\mbox{\scriptsize{eff}}}= 
({N_{\mbox{\scriptsize{grid}}}}/{N_{\mbox{\scriptsize{eff}}}})\chi^2$ 
is approximately distributed as a $\chi^2$\ with 
$N_{\mbox{\scriptsize{dof}}} = N_{\mbox{\scriptsize{eff}}}-1$\ degrees of freedom
and can be applied to infer the values of $A$, $B$ and their errors.

\begin{figure}[b]
\begin{center}
\scalebox{0.40}{\includegraphics[23,118][589,672]{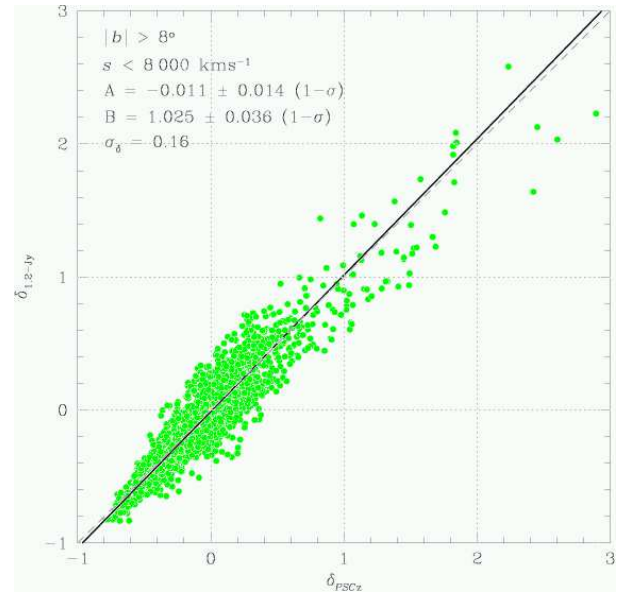}}
\caption{ \jy~vs. \pscz~overdensity. Only 1883 randomly selected gridpoints
with $|b|> 8^\circ$\ and within $s=8\,000$\kms~are shown in the scatter plot. 
The continuous line represents the linear best fit characterized by the zeropoint
$A$ and slope $B$ indicated in the panel.}
\label{fig:iras_dens_sgp}
\end{center}
\end{figure}

Fig.~(\ref{fig:iras_dens_sgp}) shows the $\delta_{1.2}$ vs. $\delta_{0.6}$ scatter
diagram obtained using all $112\,552$ gridpoints with $s < 8\,000$\kms~and 
$|b| > 8^o$. 
Within these limits the estimated  overdensities  
are little affected by  uncertainties in the filling-in procedure, and 
we expect that shot-noise dominates the total error budget. 
To estimate the shot-noise errors affecting both \pscz~and \jy~overdensity fields 
we have generated 100 bootstrap realizations of the observed \pscz~and \jy~surveys.
These are obtained by replacing each galaxy (including the `synthetic' galaxies used
to fill-in the unobserved regions) with a number of objects drawn 
from a Poisson deviate with mean unity. Shot noise errors in  $\delta$ 
at the generic gridpoint 
were set equal to the dispersion over the 100 realizations.
The solid line in the plot is the best linear fit 
obtained from the minimization of $\chi^2_ {\mbox{\scriptsize{eff}}}$.
As shown in Tab.~(\ref{t:comp}), the slope of the
line, $B = 1.025 \pm 0.036$, is consistent with unity, indicating 
no systematic mismatch. 
On the other hand  the zero-point value $A=-0.011\pm 0.014$
shows no systematic difference in the mean density of the two \iras~catalogs.

 The scatter around the best--fitting line,
$\sigma_{\delta} = 0.16$ is similar to average shot-noise error 
in the \jy~density field ($\sigma _{\mbox{\scriptsize{SN}}}^{\delta_{1.2}}= 0.15$).
The parameter $S= \chi^{\mbox{\scriptsize{eff}}}/N_{\mbox{\scriptsize{dof}}} = 0.78$ 
indicates  that bootstrap errors slightly overestimate the true uncertainties.

\subsection{Radial Density Contrast}\label{section:radialdensity5}

\begin{figure}[b]
\begin{center}
\scalebox{0.43}{\includegraphics[34,172][588,680]{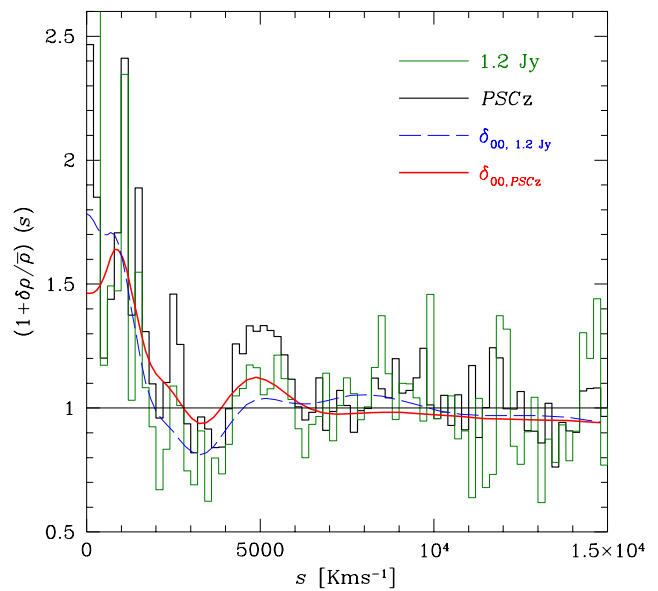}}
\caption{
Differential radial density contrast for \pscz~(thick line histogram) and \jy~galaxies 
(thin line histogram). The monopole of the \pscz~and \jy~density fields are 
shown with smooth solid and dashed lines, respectively.}
\label{fig:irasmonopole_dens}
\end{center}
\end{figure}

An alternative way to characterize possible discrepancies between the 
two IRAS density fields is that of computing the differential radial density contrast 
in shells with identical thickness $\Delta s$
for both galaxy distributions:

\begin{equation}
\frac{\rho(s)}{\bar{\rho}} =\frac{1}{V \bar{n}}\sum_{i=1}^{N_{p}} \frac{
W(s,s_i,\Delta s)
}{\phi(s_i)},
\label{eq:shellfracdensity}
\end{equation}
where the window associated to the spherical shell $W_{\delta \rho/{\bar{\rho}}}(s,s_i,$  $\Delta s)$
is the Heaviside function $H(s+\Delta
s/2-s_i)H(s_i+\Delta s/2-s)$, $s_i$ is the redshift of the object $i$,
$V$ is the volume used to compute the mean number density
${\bar{n}}$ and  the sum runs over the $N_p$ galaxies of the catalog.
Fig.~(\ref{fig:irasmonopole_dens}) shows the quantity
$\rho(s)/\bar{\rho}$ for \jy~(light, dotted histogram)
and \pscz~(thick, continuous histogram) galaxies
within  $s=15\,000$\kms, computed in  
in spherical shells of thickness  $\Delta s= 200$\kms. 
Although the general shapes of two radial density contrasts
are similar (with a bump at $s \sim 1\,100$\kms~due to
Virgo and Fornax clusters, a dip at $s\sim
3\,500$\kms~induced by the Sculptor Void and a second prominent 
peak at $s\sim 5\,000$ \kms, at the same redshift as
the Perseus-Pisces supercluster and the Great Attractor), a discrepancy
between the two curves exists in the range 
$ 2\,000\la s \la 6\,000$\kms~where the \pscz~density 
is larger than the \jy~one.

The two smoothed curves in Fig.~(\ref{fig:irasmonopole_dens}) 
show the monopole terms of the two overdensity fields $\delta_{l=0,m=0}$
as a function of radius, obtained from the spherical harmonics
decomposition. Both  curves interpolate the
histograms rather well (hence  showing the adequacy of our spherical harmonics
analysis) and confirm the existence of a \pscz~vs.~\jy~density mismatch 
between the two catalogs.

\section{The \iras~\pscz~and \jy~Peculiar Velocity Fields}\label{section:radialvelocitymaps} 

To perform a similar analysis on the  \jy~and \pscz~model 
velocity fields we have integrated Eqn.~(\ref{eq:ndchapter5})
using $\beta =0.6$, a value consistent with recent
determinations based on several velocity-velocity comparisons
(e.g. Nusser \etal~2000)\nocite{Nusser:2000},
a recent density-density analysis (Zaroubi \etal~2002) and the study of 
mean relative velocities of galaxy pairs (Juszkiewicz \etal~2000\nocite{Juszkiewicz:2000}).

\subsection{Radial Peculiar Velocity Maps}
\label{subsction:radial_velocity_maps}

To obtain maps of the radial velocity field we have computed the 
gradient of the velocity potential projected along the line of sight:
$u(\svec)= \hat{\svec} \cdot \nabla \Phi$ at the same gridpoint positions 
used for the analysis of the density field.

\begin{figure}[t]
\begin{center}
\scalebox{0.40}{\includegraphics[31,313][569,580]{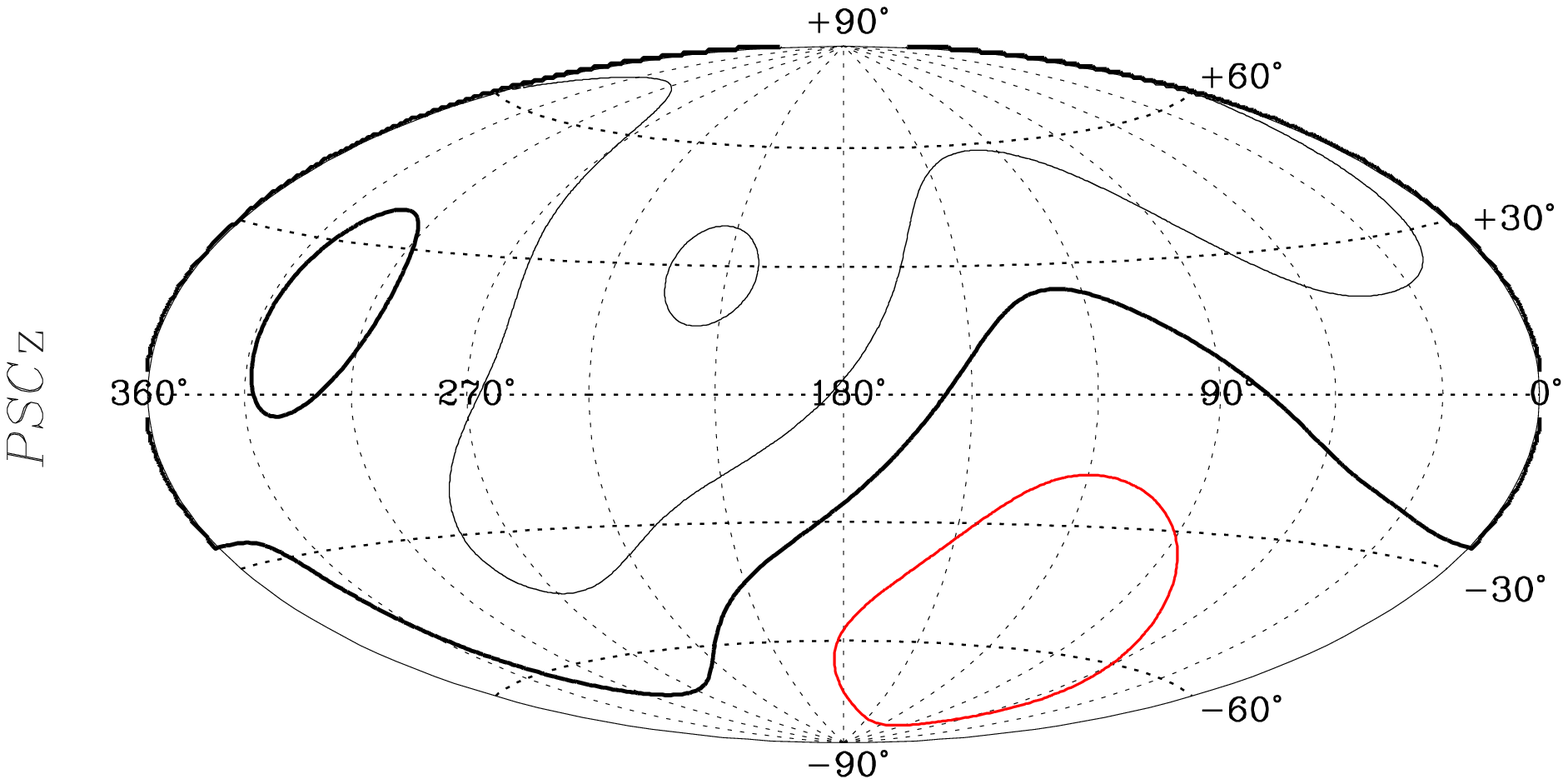}}
\scalebox{0.40}{\includegraphics[31,313][569,580]{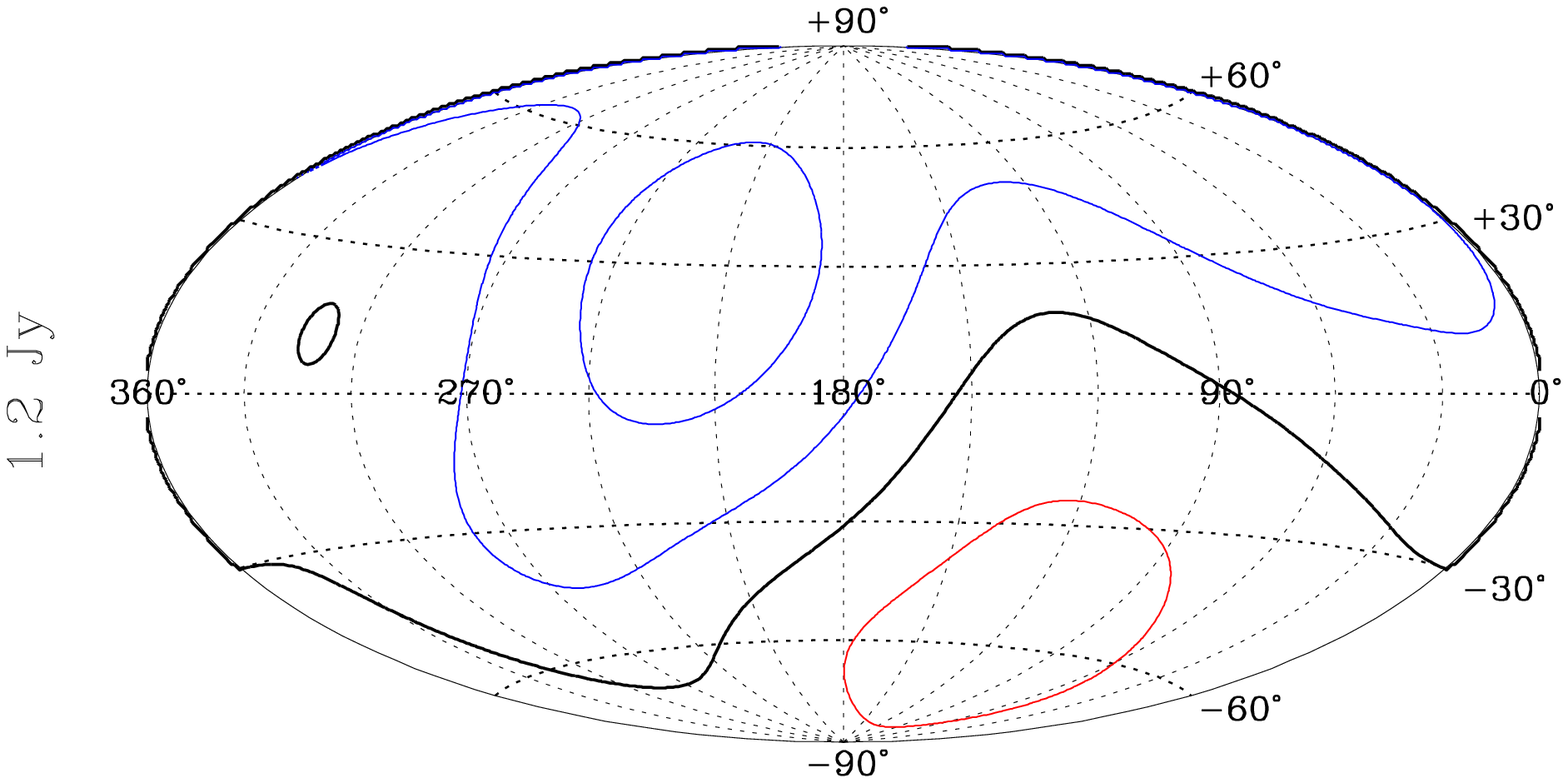}}
\scalebox{0.40}{\includegraphics[31,263][569,580]{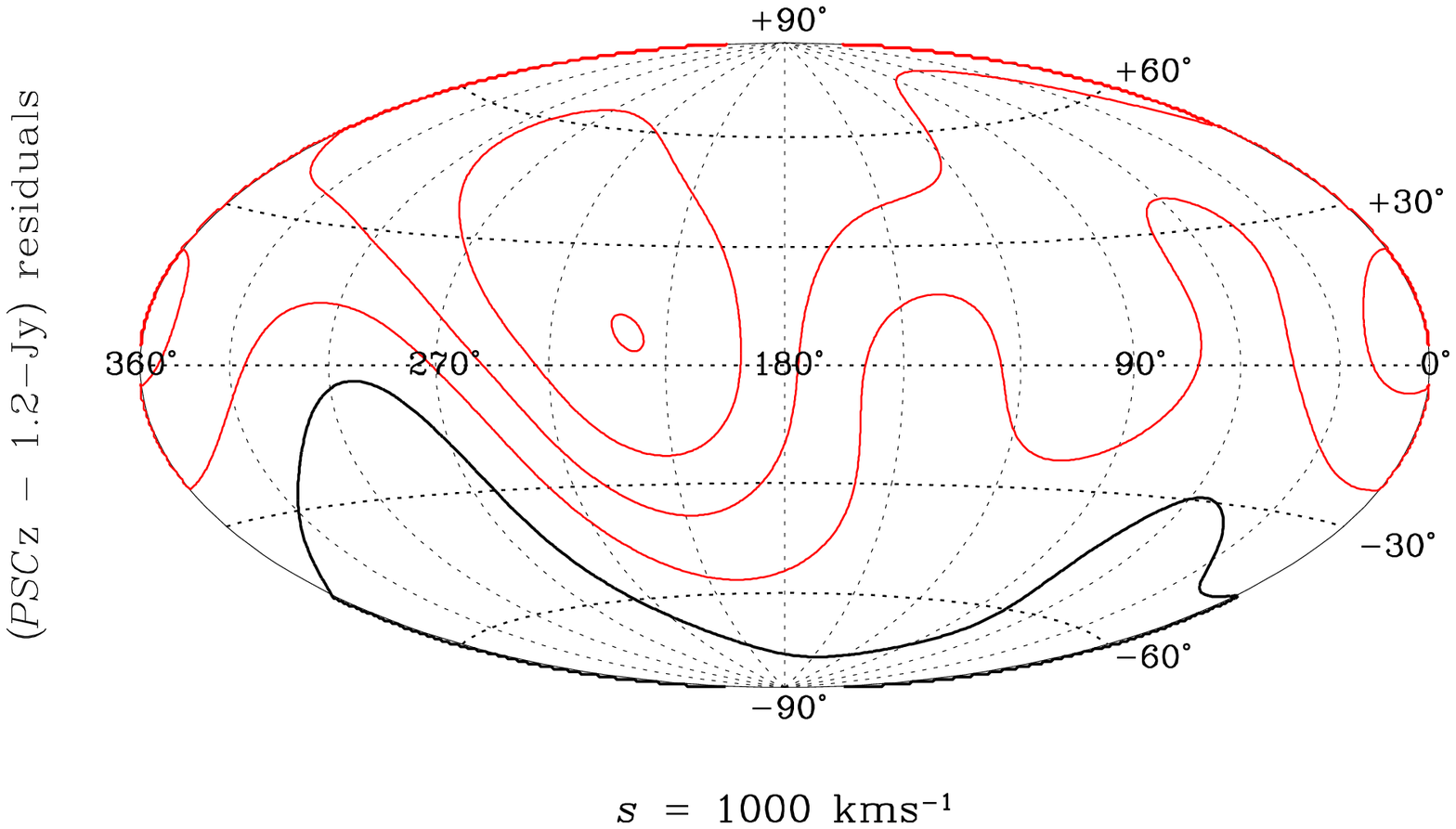}}
\caption{\pscz~(top), \jy~(middle) and residual (bottom)
radial velocity fields in the spherical shell centred  
at $s = 1\,000$\ \kms. 
The thick-line contour indicates $\delta u= 0$\ or $\Delta u_{\mbox{\scriptsize{res}}} = 0$.
The light- and dark-gray contours show regions characterized by outflows or infall 
motions. In the top and middle panels
the contours are equally--spaced by $\Delta u\ = 200$ \kms. 
The bottom panel shows  the velocity residuals in steps of 
$\Delta \delta_{\mbox{\scriptsize{res}}} = 25$ \kms.}
\label{fig:irasvel1000}
\end{center}
\end{figure}

\begin{figure}[t]
\begin{center}
\scalebox{0.40}{\includegraphics[31,313][569,580]{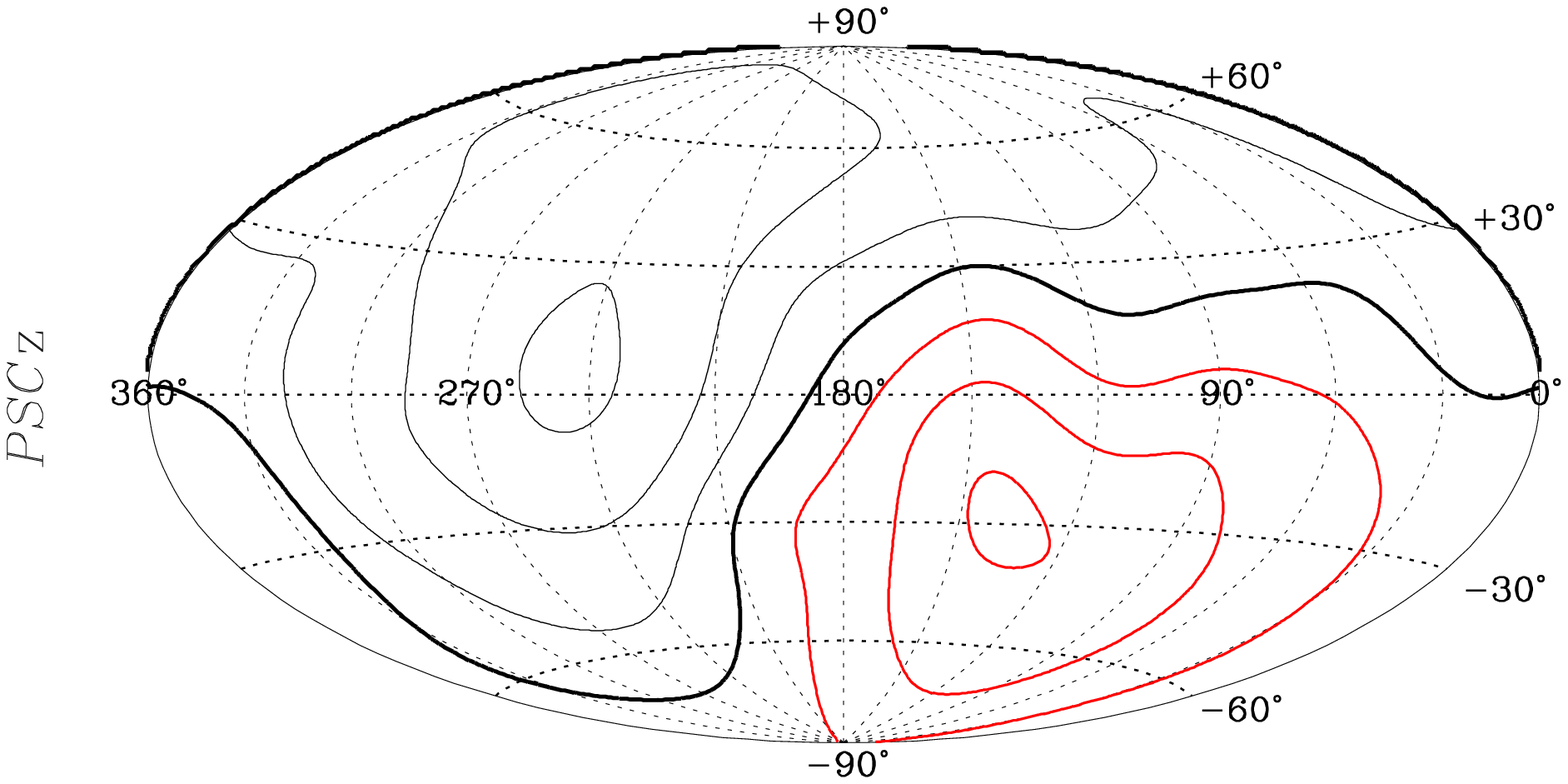}}
\scalebox{0.40}{\includegraphics[31,313][569,580]{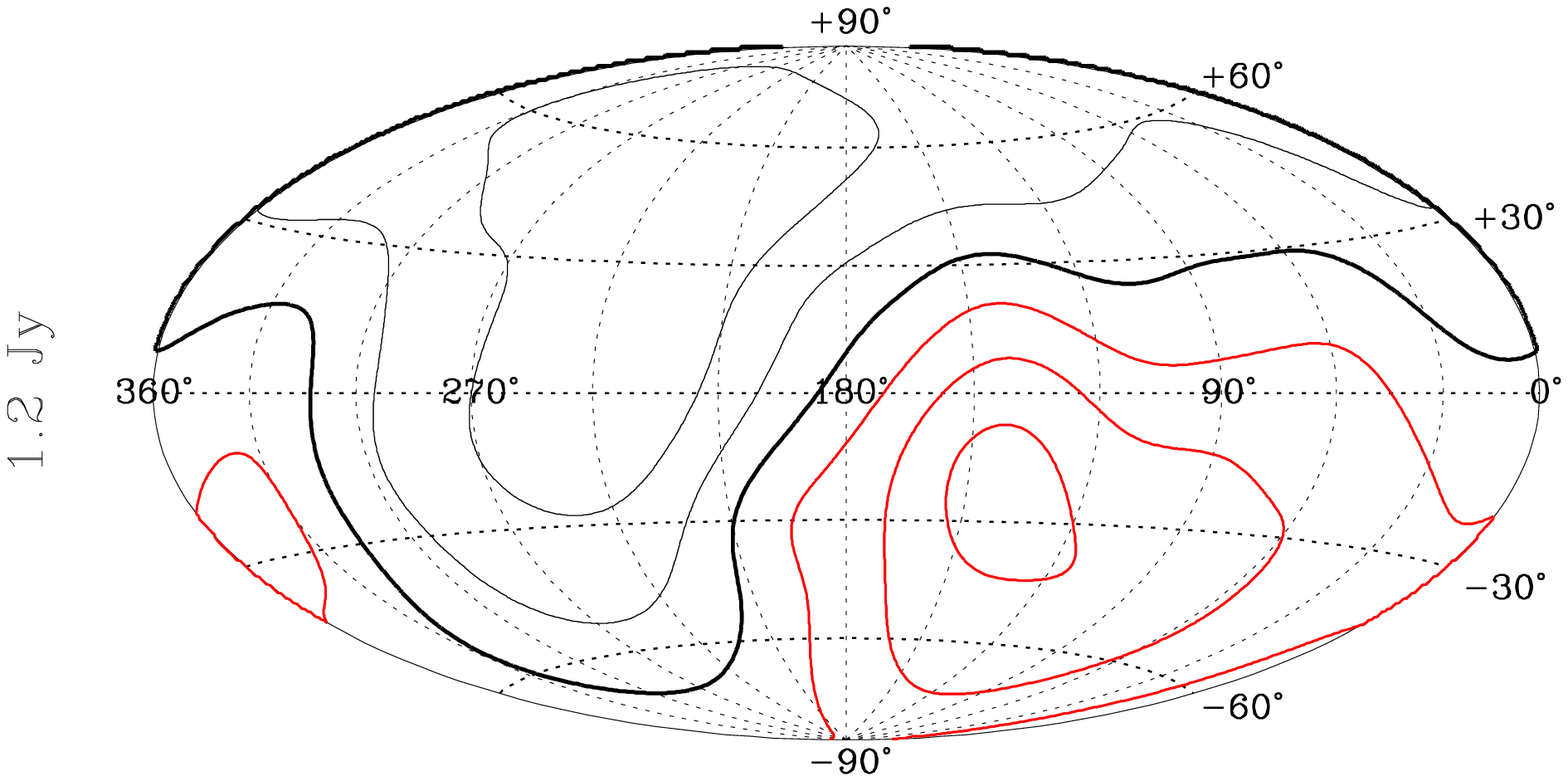}}
\scalebox{0.40}{\includegraphics[31,263][569,580]{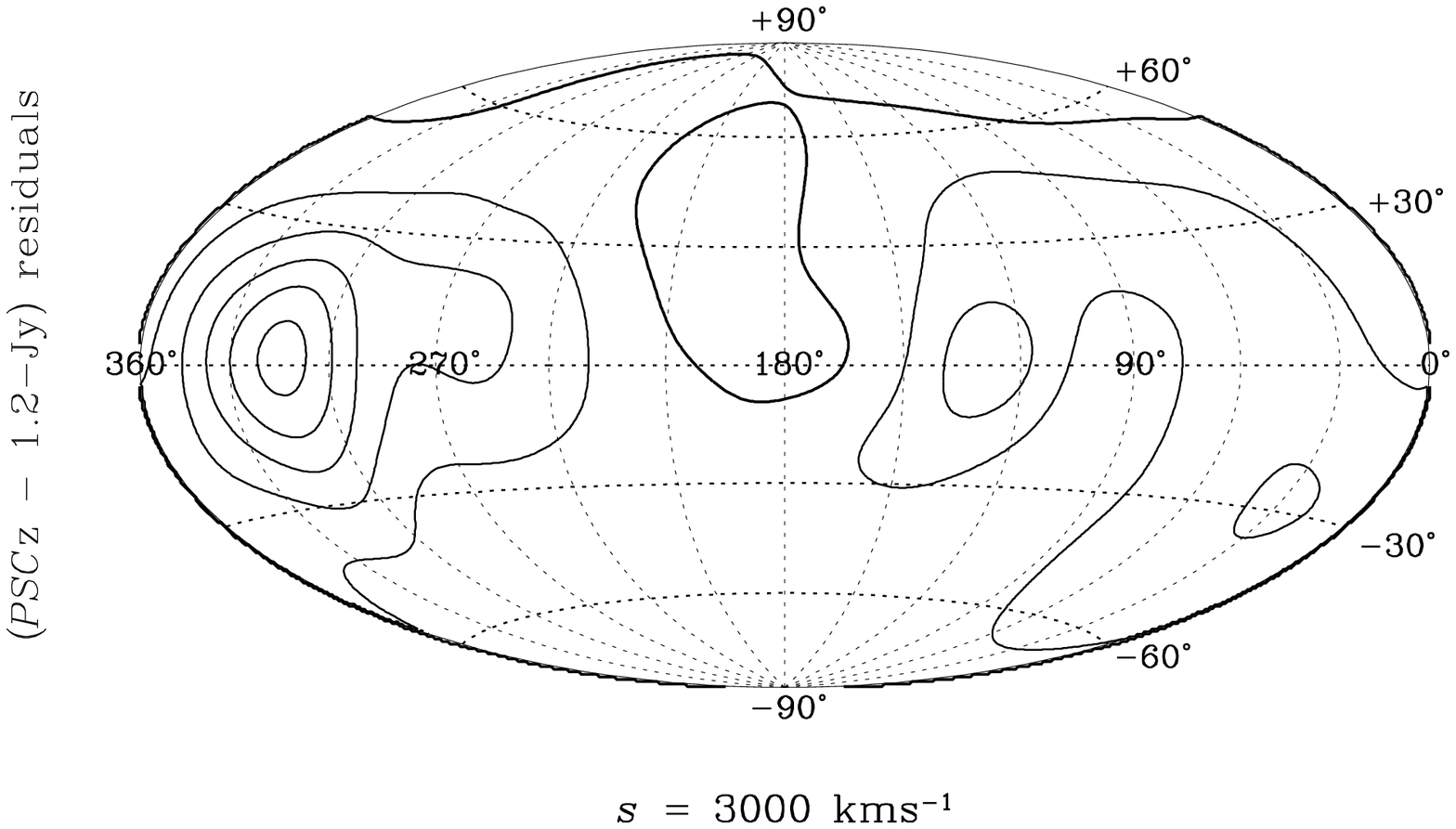}}
\caption{Same as in Fig.~(\ref{fig:irasvel1000}). The shell is now centred 
at $s = 3\,000$\kms and the contour spacings are $\Delta u\ = 200$ \kms
and $\Delta u_{\mbox{\scriptsize{res}}} = 50$ \kms}  
\label{fig:irasvel3000}
\end{center}
\end{figure}

\begin{figure}[t]
\begin{center}
\scalebox{0.40}{\includegraphics[31,313][569,580]{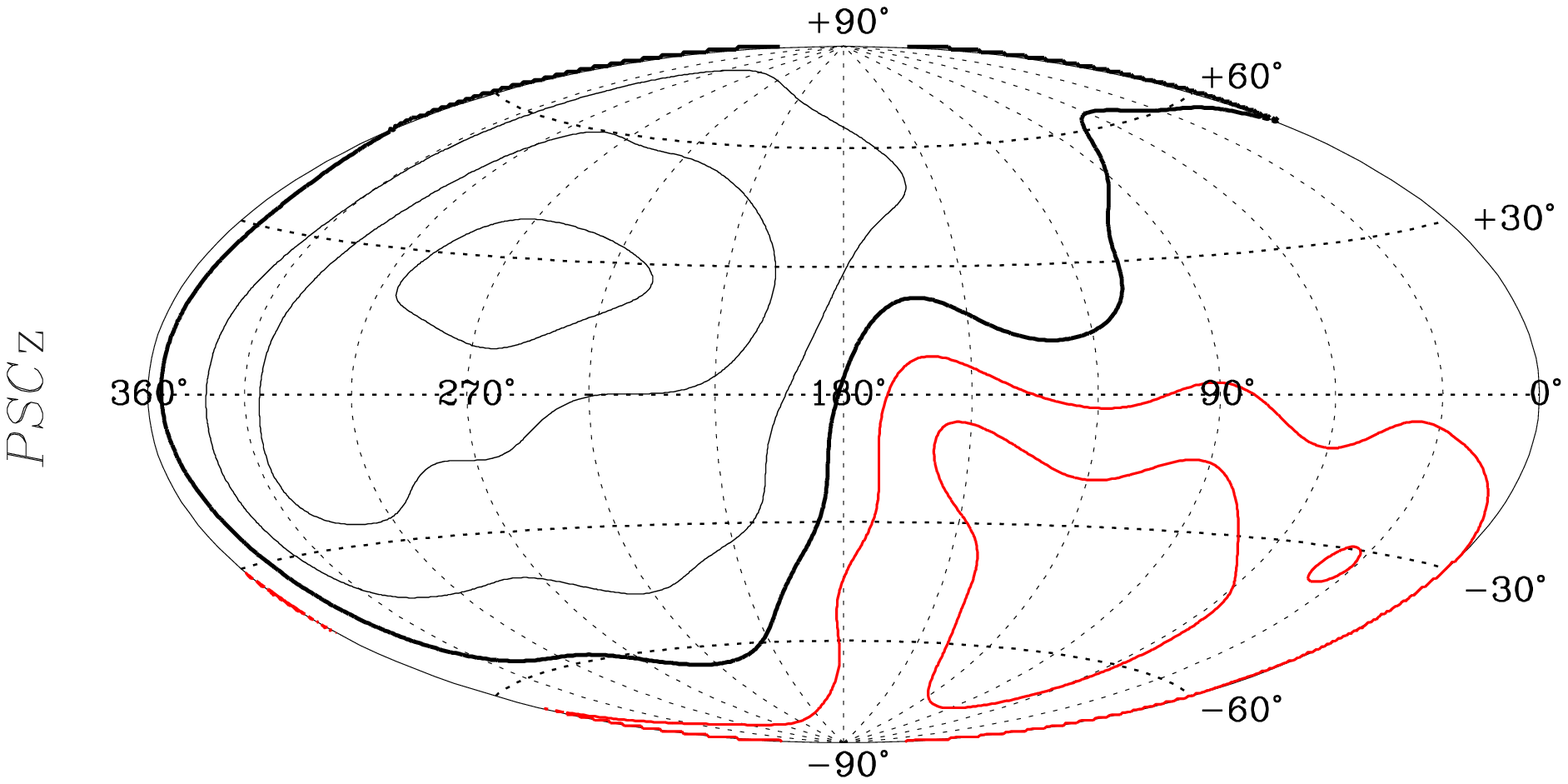}}
\scalebox{0.40}{\includegraphics[31,313][569,580]{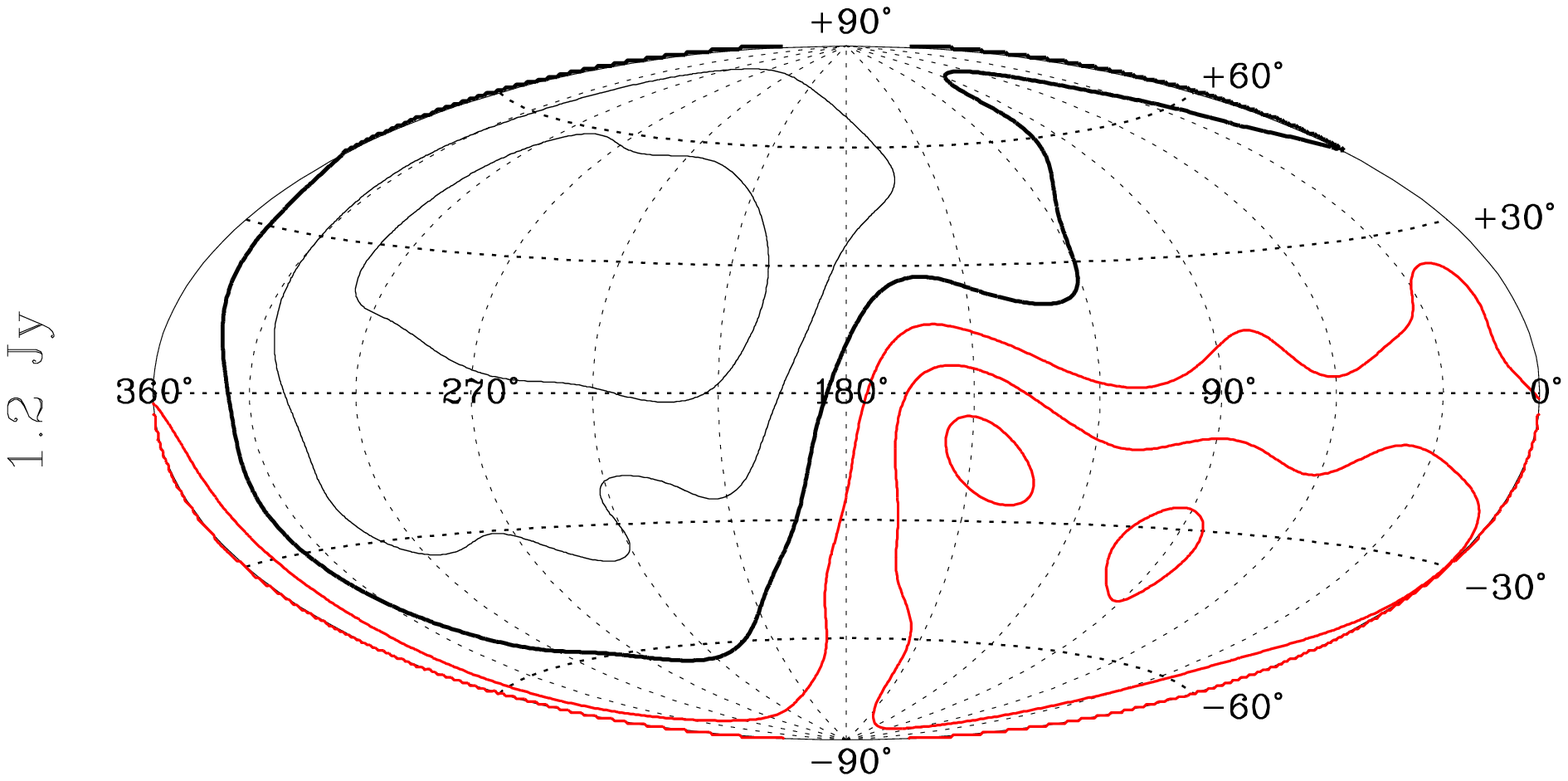}}
\scalebox{0.40}{\includegraphics[31,263][569,580]{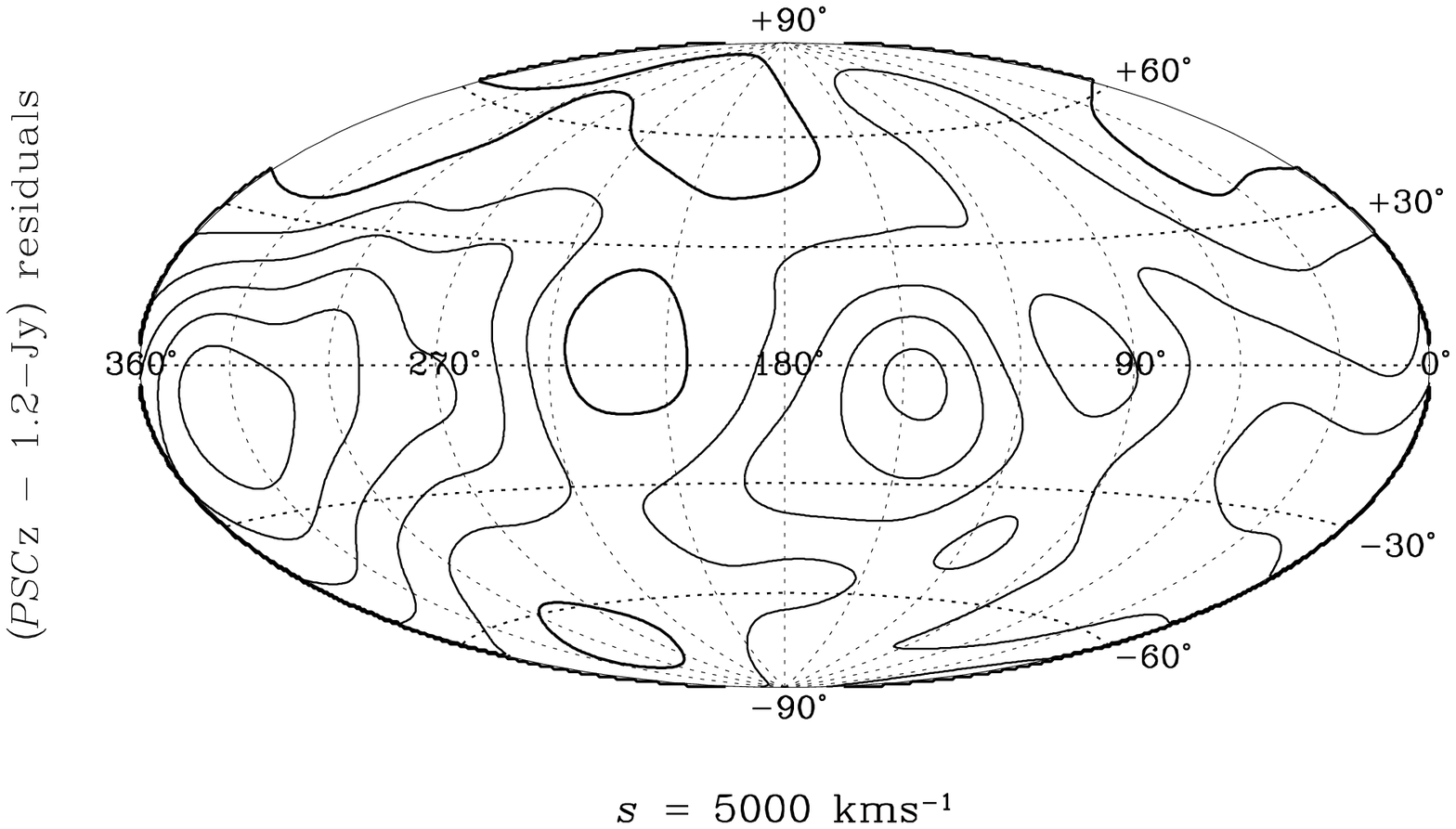}}
\caption{Same as in Fig.~(\ref{fig:irasvel1000}). The shell is now centred 
at $s = 5\,000$\ \kms and the contour spacings are $\Delta u\ = 200$ \kms
and $\Delta u_{\mbox{\scriptsize{res}}} = 50$ \kms.}
\label{fig:irasvel5000}
\end{center}
\end{figure}

The top end mid panels of Figs.~(\ref{fig:irasvel1000}-\ref{fig:irasvel5000}) 
are similar to those of figs.~\ref{fig:dens1000}--\ref{fig:dens5000}
and show the radial velocity maps in the same  shells.  The bottom panels show the
maps of the radial velocity residuals $u_{\mbox{\scriptsize{res}}}(s)\equiv u_{0.6}(s)-u_{1.2}(s)$.
The thick line represents the zero radial velocity (residual) contour.
The velocity (residual) contour spacing $\Delta u$ 
($\Delta u_{\mbox{\scriptsize{res}}}$) varies with the redshift and 
is indicated in the Figure captions.

The main feature seen at all redshifts in both \jy~and \pscz~radial velocity maps
is the dipole pattern resulting from the LG reflex motion toward the CMB apex
($l\approx 276^\circ, b\approx 30^\circ$, Kogut \etal 1993)\nocite{Kogut:1993}.

Local deviations from this pattern arise from peculiar motions, like the infall
onto Perseus-Pisces superclusters at $(s \approx 3000$ \kms,$~ l\approx 120^{\circ},~ b \approx
-40^{\circ})$. The largest velocity residuals are located near the galactic plane,
but their contours are broader due to
 the intrinsic non-locality of the peculiar velocity field.
The most striking feature is perhaps the fact that velocity residuals, that are
positive across a large fraction of the sky in the first shell, becomes mostly negative in the 
remaining two shells, suggesting possible systematic errors in either \pscz~or 
\jy~velocity models.


\subsection{Velocity Residuals}\label{section:vresiduals}

To perform a more quantitative comparison between the two 
radial velocity fields we have repeated the analysis of 
\S~\ref{section:dresiduals} and performed a linear regression
between $u_{1.2}$ and $u_{0.6}$ by minimizing 

\begin{equation}
\chi^2=\sum_{i=1}^{N_{\mbox{\scriptsize{grid}}}} 
\frac 	{ \left (u_{1.2}-A-B\sigma_{u,0.6}\right )^2 } 
	{ \left (u_{1.2}^2 + B^2 \sigma_{u,0.6}^2 \right )},
\end{equation}
where $N_{\mbox{\scriptsize{grid}}}$ runs over the same points
used for the density-density comparison.
Linear velocities are obtained by integrating the mass density fields over 
large scales causing them to be correlated on scales larger than the 
radius of the Gaussian kernel. This means that the number
of  independent gridpoint $N_{eff}$~is smaller than
the estimate given by Eqn.~(\ref{eq:neff}).
Intrinsic large scale smoothing also affects the velocity error estimate,
$\sigma_u$ since uncertainties in modeling the galaxy distribution
in the unobserved areas now add to shot noise errors.

The shot noise errors have been computed using the
same bootstrap resampling technique  described in \S~\ref{section:dresiduals}.
Uncertainties in the filling-in procedure have been evaluated from 20 
independent  \pscz~and \jy~mock catalogs
extracted from the  $N$-body simulations performed by Cole 
\etal\ (1997)\nocite{Cole:1997} simulating a volume of
$\approx 170 \hmpc$ in an Einstein-de Sitter 
universe with a non-vanishing ($\Omega_{\Lambda}= 0.7$)
cosmological constant and density  fluctuations
normalized to the observed cluster abundance
(Eke, Cole \& Frenk 1996 \nocite{Eke:1996}).
Model velocity fields from each IRAS mock catalog were compared with
velocities obtained from ideal, all-sky IRAS mock catalogs. The variance
over 20 mocks at each gridpoint quantifies the uncertainties in the filling procedure.
Total uncertainties were obtained by adding in quadrature the two errors.

The results of the linear regression are summarized in Tab.~(\ref{t:comp})
which shows the  slope of the best fitting line  $B= 1.040\pm 0.019$
and the value of the zero-point, $A= 60.0 \pm 5.9$\kms.
While the slope is still consistent with unity, at the 2.1-$\sigma$ confidence level, 
the zero point is significantly different
from zero, hence corroborating the evidence
for a mismatch in the average densities
of \pscz~and \jy~catalogs.
The dispersion around the fit is, $\sigma _{u} = 88$\kms\, is similar 
to the typical velocity error computed from the IRAS mocks and the bootstrap 
resampling catalogs 
($\sigma _{u_{0.6}} = 71$\kms~and $\sigma _{u_{1.2}} = 140$\kms). 
The parameter 
$\chi^{\mbox{\scriptsize{eff}}}/N_{\mbox{\scriptsize{dof}}}$  is less than unity 
which either indicates that errors are overestimated or that the value
$N_{eff}$, taken from Eqn.~(\ref{eq:neff}), does significantly overestimate the 
true number of degrees of freedom.
\begin{figure}[t]
\begin{center}
\scalebox{0.40}{\includegraphics[23,118][606,672]{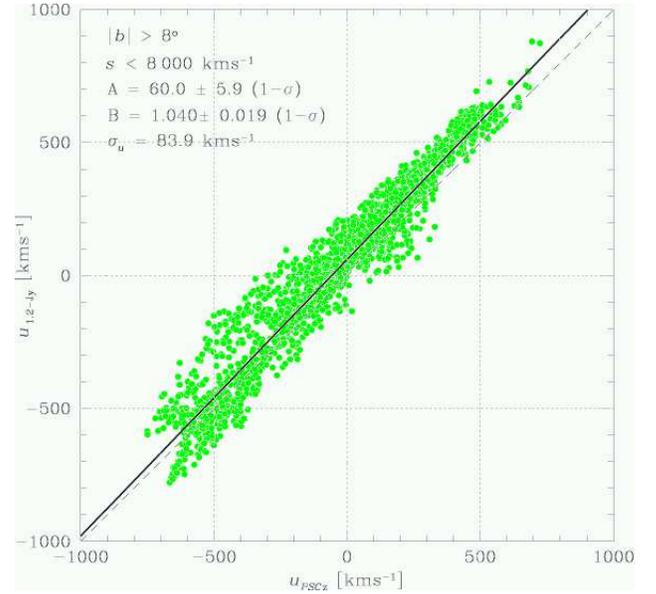}}
\caption{\jy~vs. \pscz~radial velocities computed at 1883
randomly selected gridpoint positions 
with $|b|> 8^\circ$ and  $s\le 8\,000$\kms~inferred 
radial velocity field (x-axis) versus that inferred from the \jy~survey (y-axis).
The parameters
of the best fitting line $(A,B)$\ the scatter $\sigma_{u}$ are indicated.}
\label{fig:irasmonopole_vel}
\end{center}
\end{figure}

\subsection{Velocity Multipoles}\label{section:multipoles5}

The spherical harmonics decomposition allows us to better investigate and characterize
possible discrepancies between \jy~and \pscz~velocity fields.
Here we are only interested in the first three $(l=0,1,2)$ components, 
corresponding to the monopole, dipole and quadrupole terms, which have 
already been investigated in previous analyses (e.g. BDSLS).
It is worth stressing that the monopole and dipole components at a given 
redshift are only sensitive to the mass distribution within the corresponding 
distance and thus can be directly related to the structures within the
volume considered, while the quadrupole components 
are also sensitive to the mass distribution beyond such a distance 
[see Eqn.~(9) in Nusser \& Davis 1994].

\begin{figure}[b]
\begin{center}
\scalebox{0.43}{\includegraphics[20,172][576,707]{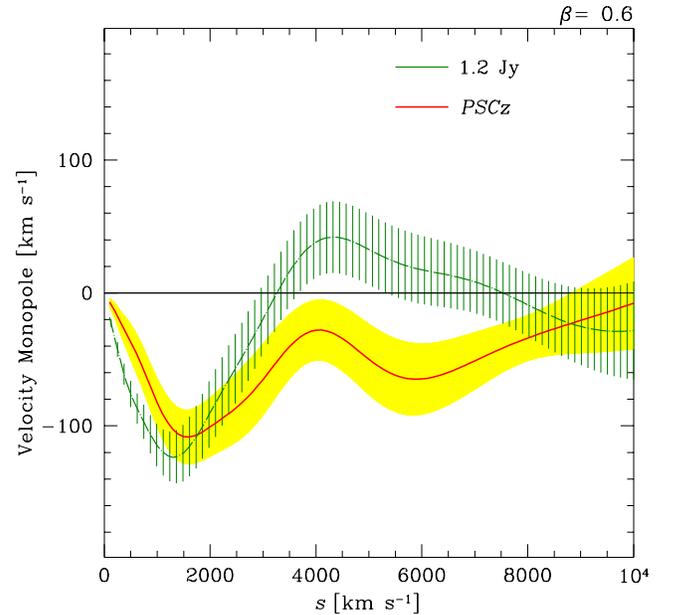}}
\caption{Monopole coefficients of \jy~(dashed)
and \pscz~(continuous) radial velocity fields.
$1-\sigma$ errors are represented by the hatched (shaded) areas.}
\label{fig:irasmonopole}
\end{center}
\end{figure}

The monopole terms of \pscz~(continuous line)
and \jy~(dashed line) velocity fields are plotted in 
Fig.~(\ref{fig:irasmonopole}) 
as a function of the redshift.
The hatched and dashed regions represent the 1-$\sigma$ uncertainty
strip, computed from the  mock catalogs and the bootstrap resampling procedure.

Both velocity monopoles display the same general features, as expected 
given the similarities between the  density monopoles $\delta_{00}(s)$
shown in Fig.~(\ref{fig:irasmonopole_dens}).
There is a significant radial inflow  
around $s = $ 1\,800\kms~corresponding to the infall
motion toward the Virgo-Fornax complex, an outflow
in the range $1\,800 \la s \la  4\,000$ \kms~due to the 
presence of the local void and another infall beyond
this radius, due to 
the Hercules, Hydra, Centaurus and Perseus-Pisces superclusters.
Differences between the two velocity monopoles  are detected
with a significance level larger than 1-$\sigma$
at very small radii ($s < 1200 $\kms) and in the range 
$3500 \ $\kms$ <s< 7500 $\kms.

\begin{figure}[t]
\begin{center}
\scalebox{0.50}{\includegraphics[31,273][520,732]{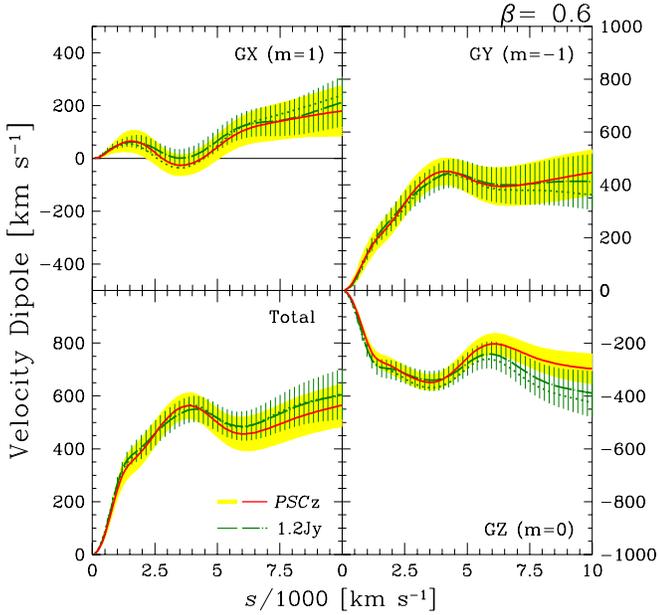}}
\caption{ The velocity dipole coefficients, $u_{1m}(s)$\{$m=-1,0,1$\}, inferred from 
\pscz~(continuous) and \jy~(dashed) and their associated 1-$\sigma$ uncertainties.
The three Galactic Cartesian components $GX$, $GY$, $GZ$
are shown in the top-left, top-right and bottom-right panel, respectively. 
The total amplitude is shown in the bottom-left panel.
The dotted line refers to the dipole of \pscz~galaxies 
with flux $f_{\mbox{\scriptsize{$60\mu m$}}} \ge 1.2$~Jy.}
\label{fig:irasdipoles}
\end{center}
\end{figure}

Fig.~(\ref{fig:irasdipoles}) shows the total amplitude (lower left panel)
and the three Cartesian components (remaining panels)
of the two IRAS velocity dipoles.
The total amplitude of the \pscz~dipole (continuous line) 
is systematically smaller that the \jy~one (dashed line)
but the discrepancy is well within the expected errors,
in  agreement with previous studies (e.g. Schmoldt \etal~1999).

\begin{figure}[t]
\begin{center}
\scalebox{0.50}{\includegraphics[31,69][511,732]{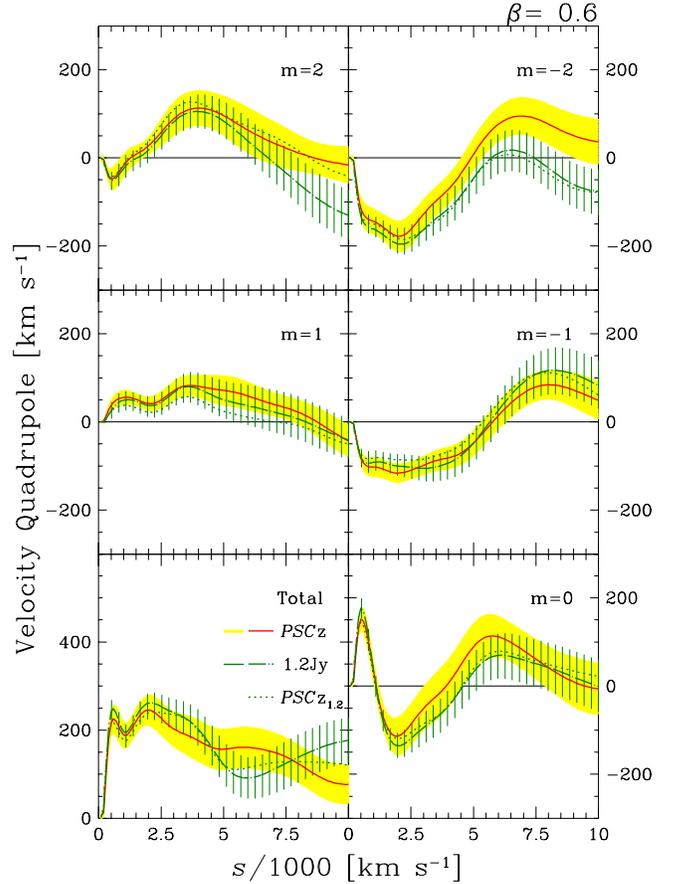}}
\caption{Quadrupole coefficients $u_{\,2m}(s)$\{$m=-2,...,2$\} derived from 
\pscz~(solid line) and \jy~(dashed) 
model velocity fields. Shaded regions indicate the $1-\sigma$\  uncertainty
regions. 
The dotted line represents the quadrupole components for \pscz~galaxies 
with flux $f_{\mbox{\scriptsize{$60\mu m$}}} \ge 1.2$~Jy.}
\label{fig:irasquadrupole}
\end{center}
\end{figure}

Finally, Fig.~(\ref{fig:irasquadrupole}) shows the five components of the 
\pscz~(continuous line) and \jy~(dashed line) velocity quadrupoles,
$u_{\,2m}(s)$ (where $m=-2,...,2$) along with total amplitude (bottom-left panel).
As for the dipole case, the two sets of quadrupole components  
agree to within 1-$\sigma$ apart from a small disagreement
in the the $m=+2$ and $m=-2$ components beyond $s=7000$ \kms.
Indeed, beyond $s\approx 6\,000$\kms~the amplitude of the \pscz~quadrupole 
decreases, while the \jy~one  increases monotonically.

It is worth noticing that our \jy~velocity multipoles agree with those 
computed by  BDSLS except that their multipole components
exhibit sharper features at small radii ($s < 500$ \kms ).
This mismatch derives from our use of a Gaussian smoothing
kernel of radius $330$\kms~which is 
somewhat larger than that applied by BDSLS.

\section{Discussion.}
\label{section:discussion5}

Perhaps the most surprising result of our analysis is the
disagreement between the \pscz~and \jy~velocity monopoles in the
$1\,800 \la s \la  4\,000$ \kms~range.
Before investigating the possible causes of this mismatch, it is
worth noticing that none of our IRAS monopoles is consistent with
an  outward flow of magnitude $ \approx 400$ \kms~at a distance 
of $ 7\,000$ \kms~like the one detected by Zehavi \etal~(1998)\nocite{Zehavi:1998}
using a sample of 44 Type Ia
supernovae. That result is also at variance with the velocity model
obtained from the distribution of ORS galaxies obtained by BDSLS 
which means that, if confirmed, it would be difficult to reconcile 
with the Gravitational Instability scenario.

A possible clue to understand the origin of the monopole mismatch 
is provided by the analysis of BDSLS that returns  a velocity monopole 
somewhere between the \pscz~and \jy~ones in the range of interest.
This seems to suggest that our monopole discrepancy originates 
from systematic errors in one of the IRAS catalogs, possibly 
some incompleteness of the largest catalog at faint fluxes.

\begin{figure}[t]
\begin{center}
\scalebox{0.43}{\includegraphics[20,172][576,706]{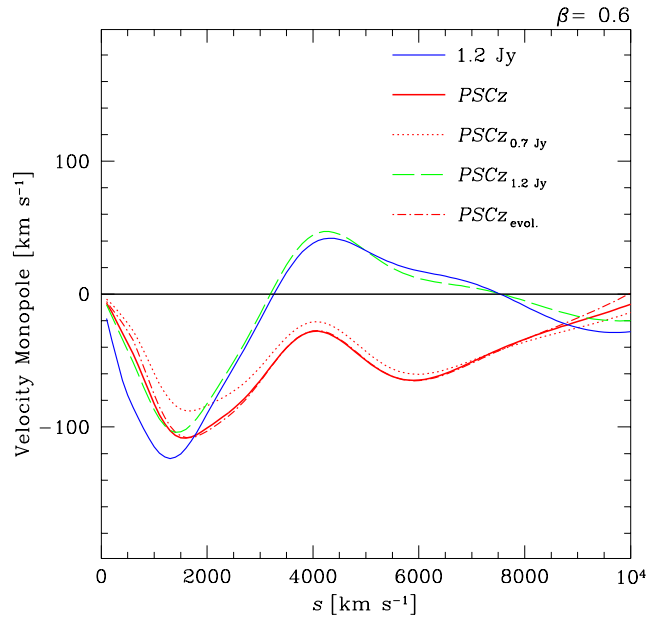}}
\caption{Velocity monopoles for the \jy~(dark gray line)
\pscz~(light-gray), \pscz$_{0.7}$\ (dot-dashed),
 \pscz$_{1.2}$\ (long-dashed) \pscz$_{\mbox{\scriptsize{evol}}}$ 
(dotted) samples. }
\label{fig:iras_monopoles_fle}
\end{center}
\end{figure}

The results of Tadros \etal~(1999)\nocite{Tadros:1999}
seem  to confirm that \pscz~catalog might be 
statistically incomplete at fluxes
$f_{\mbox{\scriptsize{$60\mu m$}}} \le 0.7$~Jy.
To test whether this is indeed the case
we have extracted a brighter \pscz~subsample (called \pscz$_{0.7}$)
by discarding all object with $f_{\mbox{\scriptsize{$60\mu m$}}} \le 0.7$~Jy
and have repeated our spherical harmonics decomposition analysis.
The results, shown in  Fig.~(\ref{fig:iras_monopoles_fle})
indicate that the \pscz$_{0.7}$ velocity monopole (dotted line)
basically coincide with that of the complete sample.
It is only when discarding galaxies with fluxes 
smaller than  1.2 Jy that the  velocity  monopole 
(dashed line) agrees with the \jy~one.

Another possibility of understanding the monopole mismatch is that 
of invoking a strong evolution of IRAS galaxies as a function 
of the redshift, which  we have neglected in our calculations so far. 
Roughly speaking evolution can be distinguish into two different types.
Pure density evolution changes only the normalization of the 
luminosity function while its shape is invariant with redshift. 
Pure luminosity evolution on the other hand describes a possible 
change of the intrinsic luminosities of the sources but not their 
number density.
Both forms of evolution occur simultaneously in the Universe. 
For instance, galaxy merging reduces the number density of galaxies  
causing thus evolution. Besides, such events also cause 
different luminosities of the final systems and hence leading to 
luminosity evolution. Other physical mechanism for galaxy evolution
can be devised. Some evidences for a sizable evolution of \iras~galaxies 
have been reported by a number of authors,
although there seem to be much controversy about the amplitude of the
effect (see Springel 1996\nocite{Springel:1996} and references therein). 

Canavezes \etal~(1998)\nocite{Canavezes:1998} expressed 
evolution in terms of a generalization of the selection function 
introduced by Yahil \etal~(1991) given by

\begin{equation}
\phi(s)_{\mbox{\scriptsize{evol}}} = \left \{ \begin{array}{ll}  \left ( \frac{r_s}{r} \right )^\alpha \left ( \frac{ r_{\star}^\gamma +  r_{s}^\gamma} 
{r_{\star}^\gamma + r^\gamma}\right )^{\beta/\gamma} &\quad,~s \ge H_0 r_s,\\
 					1&\quad,~\mbox{otherwise.}\\
\end{array}
\right .
\label{eq:volker_sf}
\end{equation}
Springel (1996) showed that it indeed provides a rather accurate modeling of both \iras~data-sets.
The estimated parameters using a maximum-likelihood procedure 
are listed in Tab.~(\ref{selfpsczsub}) and 
are denoted by \pscz$_{\mbox{\scriptsize{evol}}}$. 
The effect of evolution are shown in Fig.~(\ref{fig:iras_monopoles_fle})
in which   the velocity monopole term derived from the \pscz~galaxy
distribution in which evolutionary effects have been accounted for 
through the new selection function $\phi(s)_{\mbox{\scriptsize{evol}}}$
(dot--dashed line) is plotted against the unevolved model. 
The evolutionary effects are very minor, as was somewhat expected given 
the locality of our galaxy sample, and do not help in reducing the 
\pscz~vs. \jy~velocity monopole mismatch.

Overall, our analysis seems to indicate that the discrepancy between 
\jy~and \pscz~velocity monopoles cannot be ascribed to errors in 
the modeling of the selection function which, as we have seen, would  not 
change the monopole terms appreciably. Catalog incompleteness at very low fluxes
has to be ruled out too. We can only conclude that either the PSC$z$ incompleteness
extend at objects with fluxes comparable to 1.2 Jy or that the monopole mismatch 
does reflect a genuine difference in the average density of faint vs. bright 
IRAS galaxies at $s \le 4000$ \kms.

\begin{table}[b]
\begin{center}
\caption[]{Selection function parameters In the \iras~mock catalogs.}
\tabcolsep 1pt
\begin{tabular}{ccccc} \\\hline \hline
\\
{Mock catalog}~~~&  {~~~$\alpha$~~~}  & {~~~$\beta$~~~}  &~~~~$r_{s}$~~~& ~~~$r_{\star}$~~~ \\  
  & &   & ~$[\hm]$~ & ~$[\hm]$~ \\
\\
\hline
\\

\pscz~& 0.53 & 1.90 & 10.90 & 86.40 \\
\jy~ & 0.48 & 1.79 & 5.00 & 50.40 \\

\\
\hline \hline
\end{tabular}
\label{t:self}
\end{center}
\end{table}

A second important result found in our analysis is that
the \jy~and \pscz~dipoles agree within the 1-$\sigma$ errors.
This means that the use of the \pscz~rather than the \jy~catalog 
does not help in reducing the \markiii-IRAS dipole residuals found 
by Davis, Nusser \& Willick (1996). Indeed, both GX and GY
components of the \pscz~velocity dipole are much too large with respect to
 \markiii~ones (see Fig 15 of Davis,
Nusser \& Willick 1996). Similar results were also obtained 
from the ORS vs. \markiii~dipole velocity comparison (BDSLS).
We conclude that the discrepancy between the \markiii~and \jy~velocity 
fields cannot be alleviated by reducing the shot noise 
errors of the model velocity field.

The agreement between the \pscz, \jy~and ORS dipoles
is reassuring  since it guarantees that uncertainties
in model velocity fields arising from within the galaxy samples
that derive from incompleteness 
in one (or more) catalogs, incorrect treatment of non-linear
velocities or non-linear non-uniform biasing do not generate
systematic errors apart from a the monopole mismatch in the range
$1\,800 $\kms$\le s \le  4\,000$ \kms.

Unlike the case of monopole and dipole moments, possible 
differences between the \jy~and \pscz~quadrupole components
can also be attributed to incorrect determination and dilute sampling
of the density field beyond the sample.
The fact that \jy~and \pscz~velocity quadrupole 
agree within the expected errors suggests that differences 
in the quadrupole components of the two fields can be understood 
in terms of shot noise errors and uncertainties in filling the empty regions.

With this respect it is worth investigating whether the denser sampling of  
the \pscz~catalog at large radii allows a better modeling of the large scale contribution
to the model velocity field. 
Indeed, when comparing the model \jy~velocity field to \markiii~velocities
in the framework of the VELMOD analysis, WSDK have found that
velocity residuals exhibit a quadrupole pattern of the form
$u_{\mathcal{Q}} = \mathcal{V}_{\mathcal{Q}}\,\rvec\, \cdot \, \hat{\rvec}$, 
where $\mathcal{V}_{\mathcal{Q}}$
is a traceless, symmetric $3\times 3$ tensor.
They have attributed  this ``VELMOD quadrupole'' residuals
to uncertainties in modeling the IRAS density field beyond 
 $3\,000$ \kms. In particular they found that
the most important sources of uncertainties were the smoothing
procedure (based on a Wiener Filtering technique) and the 
shot noise errors, with the errors deriving from having ignored
mass inhomogeneities beyond $12,000$ \kms~playing a minor role.
Using \pscz~instead of \jy~catalog reduces the shot 
noise errors in the density field beyond $3000$ \kms~and thus should 
improve the agreement with the \markiii~velocities.
In other words one would expect the quadrupole 
of the residual velocity field $u^>_{0.6}-u^>_{1.2}$, where 
one only takes in account the mass beyond $3000$ \kms, to be similar
to  the ``VELMOD quadrupole'' of WSDK.

\begin{figure}[b]
\begin{center}
\scalebox{0.43}{\includegraphics[23,126][590,666]{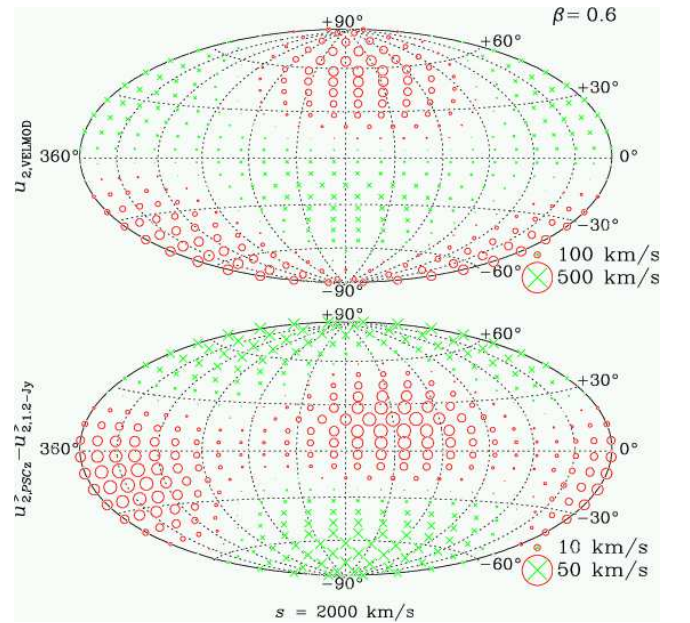}}
\caption{VELMOD (upper panel) and $u^>_{0.6}-u^>_{1.2}$ (lower panel) quadrupole residuals 
at $s = 2\,000$\kms. Stars denote outflowing from us (LG frame); circles denote infall to us.
The size of the symbol is proportional to the amplitude 
of the flow (notice the different velocity scales in the two panels).
The maximum amplitude of the ``VELMOD quadrupole'' (top panel) 
is $\approx$~181\kms, while for $u^>_{0.6}-u^>_{1.2}$ is $\approx$~23 \kms.
\label{fig:external_quad}}
\end{center}
\end{figure}

In Fig.~(\ref{fig:external_quad}) we compare the
VELMOD quadrupole at $s=2000$\kms~(rescaled from $\beta=0.492$ 
to $\beta=0.6$, upper panel) 
to the  $u^>_{0.6}-u^>_{1.2}$ quadrupole residual (lower panel). 
The VELMOD quadrupole reaches its maximum value at 
$(l\approx 165^{\circ},~b \approx 55^{\circ})$
and in the opposite direction of the sky.
The $u^>_{0.6}-u^>_{1.2}$ residual quadrupole exhibits a rather different
pattern, reaching its maximum amplitude at 
$(l\approx 156^{\circ},~b \approx -74^{\circ})$.
The value of the quadrupole components of \jy~and \pscz~as 
well as those of the VELMOD and $u^>_{0.6}-u^>_{1.2}$ quadrupole residuals
computed at a distance of 2000 \kms~
are shown in Tab.~(\ref{t:VELquad}). This comparison is at best 
qualitative since ({\it i}) VELMOD quadrupole 
is computed in real space while both IRAS quadrupoles refer to redshift 
space insted, and ({\it ii}) the smoothing scheme used in WDKS is different from 
with ours.

\begin{table}[b]
\begin{center}
\caption[]{Quadrupole Components at $2\,000$\kms}
\tabcolsep 1pt
\begin{tabular}{crrrr} \\\hline \hline
\\
{ }~~~& ~VELMOD~ & ~~~\pscz~~~\,   & ~~~~\jy~~~~ & ~{\pscz~ - \jy}~ \\
                 & [\kms]~~\,  & [\kms]\, & [\kms]~~ & [\kms]~~~~~\, \\
\\
\hline
\\
$\cal{V}_Q$ (x,x)...~~~~ &   45~~~~~~~  &   57~~~~~~ &   69~~~~~~ & -13~~~~~~~~~~~ \\
$\cal{V}_Q$ (y,y)...~~~~ &   44~~~~~~~  &  -22~~~~~~ &  -15~~~~~~ &  -7~~~~~~~~~~~ \\
$\cal{V}_Q$ (z,z)...~~~~ &  -89~~~~~~~  &  -35~~~~~~ &  -54~~~~~~ &  20~~~~~~~~~~~ \\
$\cal{V}_Q$ (x,y)...~~~~ &   18~~~~~~~  & -150~~~~~~ & -156~~~~~~ &   6~~~~~~~~~~~ \\
$\cal{V}_Q$ (x,z)...~~~~ &  138~~~~~~~  &   57~~~~~~ &   48~~~~~~ &  10~~~~~~~~~~~ \\
$\cal{V}_Q$ (y,z)...~~~~ &  -29~~~~~~~  & -106~~~~~~ & -102~~~~~~ &  -5~~~~~~~~~~~ \\
\\
\hline \hline
\end{tabular}
\label{t:VELquad}
\end{center}
\small{NOTE.--(1) We have rescaled the ``VELMOD quadrupole'' components
presented in WSDK's Tab.~(2) to the numerical value of $\beta=0.6$.}
\end{table}

\section{Conclusions.}
\label{section:conclusions5}

We have compared the model velocity and density fields of the 
\pscz~survey to those derived from the sparser \jy~sample 
in redshift space. Our model predictions are based on the 
 Nusser \& Davis (1994)\nocite{Nusser:1994} 
spherical harmonics decomposition technique
and  assumes Zeld'ovich approximation and linear biasing.

The IRAS overdensity fields are reconstructed with a smoothing proportional
to the \jy~inter-galaxy separation and  exhibit
the same general features corresponding to the known cosmic structures 
in the local universe.
The model  velocity field, computed in the LG frame, has a dipole--like
appearance, with galaxies moving away from us toward Perseus-Pisces region
and approaching the LG from the opposite direction.

The analysis of the velocity multipoles has revealed a significant discrepancy
between the \pscz~and \jy~velocity mono- poles in the range 
 $ 1\,800 \la s \la 4\,000$\kms~in which \pscz~radial velocities are 
systematically larger than \jy~ones, that cannot be explained 
by uncertainties in the IRAS selection functions or incompleteness
of the \pscz~catalog at faint fluxes.
Also, the mismatch cannot be ascribed to errors derived from
shot--noise and treatment of unobserved areas since both have been
accounted for with the help of realistic mock IRAS catalogs and 
the extensive application of bootstrap
resampling techniques to both IRAS datasets.

Both IRAS dipole and quadrupole moments are in good agreement 
within the sampled volume apart from a small mismatch in the 
$u_{10}$, $u_{2,-2}$\ and $u_{22}$ quadrupole components 
beyond $s=7\,000$\kms.
The agreement between the dipole moments and their consistency with
the velocity dipole computed by BDSLS from the ORS 
catalog  suggests that deviations 
from homogeneous linear biasing prescriptions have to be small 
and cannot be advocated to explain the aforementioned mismatch 
between the IRAS monopoles.

Another remarkable consequence of the agreement among the 
various dipoles is that the \jy-\markiii~residuals
cannot be alleviated by using the  \pscz~catalog instead of 
the \jy~or the ORS ones. In fact, this suggests that 
the origin of the \jy-\markiii~dipole resides in the 
\markiii~calibration procedure; a suggestion  corroborated by 
the fact that the dipole mismatch disappears when adopting
the alternative VELMOD procedure to calibrate -\markiii~velocities (WDSK).
The VELMOD analysis requires a sizable external quadrupole 
component within $3000$ \kms. This quadrupole accounts
for uncertainties errors in the \jy~model density field beyond that distance.
We found that using the  \pscz~sample provides part of the
required external tidal field.

As a final remark, we would like to stress that the present analysis
illustrates once more the importance of spherical harmonics decomposition
techniques to compare observed and model density and velocity fields
obtained from all-sky samples, to reveal possible data-model or model-model
inconsistencies and to assess model adequacy.
These techniques will prove very useful when the next generation of all-sky 
redshift surveys and peculiar velocity catalogs [e.g. the 6dF (\url{http:www.mso.anu.
edu.au/6dFGS}) and 2MRS (\url{http:cfa-www.harvard.edu/huchra/ 2mass}) datasets] will become available.

\section*{Ackowledgements}
We thank Adi Nusser for enlightening discussions.
LT was partly funded by FCT (Portugal) under the grants
PRAXIS XXI /BPD/16354/98, PRAXIS/C/FIS/13196/98 and POCTI/1999/
FIS/36285.
LT aknowledges the hospitality of the Aspen Center 
for Physics and the Kvali Institute for Theoretical Physics
where much of the work was completed.


\label{lastpage}
\end{document}